\documentclass[journal]{IEEEtran}
\usepackage[T1]{fontenc}
\usepackage{graphicx}
\usepackage{cite}
\usepackage{subcaption}

\usepackage{lipsum}
\usepackage{overpic}
\usepackage{amssymb}
\usepackage{amsmath}
\usepackage{amsthm}
\usepackage{array}
\usepackage{color}
\usepackage{url}

\IEEEoverridecommandlockouts

\theoremstyle{plain}

\newtheorem{definition}{Definition}
\newtheorem{lemma}{Lemma}
\newtheorem{assumption}{Assumption}
\newtheorem{corollary}{Corollary}
\newtheorem{remark}{Remark}
\newtheorem{proposition}{Proposition}

\newcommand{\mathacr}[1]{\mathsf{#1}}

\newcommand{\vect}[1]{\mathbf{#1}}

\def\diag{\mathrm{diag}}

\def\tr{\mathrm{tr}}

\def\Htran{\mbox{\tiny $\mathrm{H}$}}
\def\Ttran{\mbox{\tiny $\mathrm{T}$}}
\def\CN{\mathcal{N}_{\mathbb{C}}} 

\begin{document}

\title{Scalable Cell-Free Massive MIMO Systems}

\author{\IEEEauthorblockN{Emil Bj{\"o}rnson, \emph{Senior Member, IEEE}, Luca Sanguinetti, \emph{Senior Member, IEEE}
\thanks{\copyright 2020 IEEE. Personal use of this material is permitted. Permission from IEEE must be obtained for all other uses, in any current or future media, including reprinting/republishing this material for advertising or promotional purposes, creating new collective works, for resale or redistribution to servers or lists, or reuse of any copyrighted component of this work in other works.
\newline\indent This article was presented in part at IEEE PIMRC 2019 \cite{BjornsonPIMRC2019}.
\newline\indent E. Bj{\"o}rnson was supported by ELLIIT and the Wallenberg AI, Autonomous Systems and Software Program (WASP). L. Sanguinetti was supported by the University of Pisa under the PRA 2018-2019 Research Project CONCEPT.
\newline \indent E.~Bj\"ornson is with the Department of Electrical Engineering (ISY), Link\"{o}ping University, 58183 Link\"{o}ping, Sweden (emil.bjornson@liu.se). L.~Sanguinetti is with the University of Pisa, Dipartimento di Ingegneria dell'Informazione, 56122 Pisa, Italy (luca.sanguinetti@unipi.it).}
}}
\maketitle

\maketitle
\begin{abstract}
Imagine a coverage area with many wireless access points that cooperate to jointly serve the users, instead of creating autonomous cells. Such a cell-free network operation can potentially resolve many of the interference issues that appear in current cellular networks. This ambition was previously called Network MIMO (multiple-input multiple-output) and has recently reappeared under the name Cell-Free Massive MIMO. The main challenge is to achieve the benefits of cell-free operation in a practically feasible way, with computational complexity and fronthaul requirements that are scalable to large networks with many users. We propose a new framework for scalable Cell-Free Massive MIMO systems by exploiting the dynamic cooperation cluster concept from the Network MIMO literature. We provide a novel algorithm for joint initial access, pilot assignment, and cluster formation that is proved to be scalable. Moreover, we adapt the standard channel estimation, precoding, and combining methods to become scalable. A new uplink and downlink duality is proved and used to heuristically design the precoding vectors on the basis of the combining vectors. Interestingly, the proposed scalable precoding and combining outperform conventional maximum ratio processing and also performs closely to the best unscalable alternatives.
\end{abstract}

\begin{IEEEkeywords}
Cell-Free Massive MIMO, scalable implementation, centralized and distributed algorithms, dynamic cooperation clustering, user-centric networking, uplink-downlink duality.
\end{IEEEkeywords}

\IEEEpeerreviewmaketitle

\section{Introduction}

By transmitting a signal coherently from multiple antennas, the received power can be increased without increasing the total transmit power \cite{massivemimobook}. This is the phenomenon utilized by classic beamforming from co-located antenna arrays, but can be also utilized when transmitting coherently from multiple access points (APs) \cite{Shamai2001a}. Even if the APs have different channel gains to the user equipment (UE), the benefit of coherent transmission makes it better to divide the transmit power over multiple APs than transmitting only from the AP with the best channel. Such coherent transmission has been given many names, including Network MIMO  \cite{Venkatesan2007a}, and provides substantially higher performance than when each UE is only served by one AP \cite{Ngo2017b,Nayebi2017a,Bjornson2019c}.

The early Network MIMO papers assumed all APs have network-wide channel state information (CSI) and transmit to all UEs \cite{Venkatesan2007a,Gesbert2010a}. These are two theoretically preferable, but impractical, assumptions that lead to immense fronthaul signaling for CSI and data sharing, respectively, as well as huge computational complexity.
Fortunately, \cite{Bjornson2010c} proved that Network MIMO can operate without CSI sharing, by sacrificing the ability for the APs to jointly cancel interference.
Moreover, to limit data sharing and computational complexity, each UE can be served only by an AP subset \cite{Bjornson2013d}. Initially, a \emph{network-centric} approach was taken by dividing the APs into non-overlapping (disjoint) cooperation clusters in which the APs are sharing data (and potentially CSI) to serve only the UEs residing in the joint coverage area \cite{Marsch2008a,Zhang2009b,Huang2009b}. This approach was considered in 4G but provides small practical gains \cite{Fantini2016a}. One key reason is that many UEs will be located at the edges of the clusters and, thus, will observe substantial inter-cluster interference from the neighboring clusters \cite{Garcia2010a}.

The alternative is to take a \emph{user-centric} approach where each UE is served by the AP subset providing the best channel conditions. Since these subsets are generally different for every UE, it is not possible to divide the network into non-overlapping cooperation clusters. Instead, each AP needs to cooperate with different APs when serving different UEs, over the same time and frequency resource \cite{Tolli2008a,Bjornson2011a,Kaviani2012a}.\footnote{It has also been proposed to have non-overlapping cooperation clusters that change over time or frequency
\cite{Papadogiannis2008a,Jungnickel2014a}. One can then mitigate inter-cluster interference by scheduling each UE on its preferred cluster configuration. However, this is an inefficient solution since each UE can only be assigned to a fraction of the time-frequency resources, while MIMO systems should preferably assign all such resources to all UEs and separate UEs spatially.}
A general user-centric cooperation framework was proposed in \cite{Bjornson2011a} under the name \emph{dynamic cooperation clustering} (DCC) and was further described and analyzed in the textbook \cite{Bjornson2013d}. The word \emph{dynamic} refers to the adaptation to time-variant characteristics such as channel properties and UE locations (to name a few). The practical feasibility  of DCC was experimentally verified by the pCell technology \cite{Perlman2015a}, but the combination of Network MIMO and DCC didn't gain much interest at the time it was proposed since Massive MIMO was simultaneously conceived \cite{Marzetta2010a} and rightfully gained the spotlight.

\subsection{Cell-Free Massive MIMO and Scalability Issues}

Now that Massive MIMO is a rather mature technology \cite{massivemimobook,Sanguinetti2019a,Bjornson2019a} that has made its way into the 5G standard \cite{Parkvall2017a}, the research focus is shifting back to Network MIMO, but under the new name \emph{Cell-Free Massive MIMO} coined in \cite{Ngo2017b,Nayebi2017a}. The key novelty is the spectral efficiency (SE) analysis that features imperfect CSI, but conceptually, it is a special case of Network MIMO operating in time-division duplex (TDD) mode. In fact, it has been a step backwards in terms of implementation feasibility. Although most papers embraced the approach from \cite{Bjornson2010c} of not sharing CSI between the cooperating APs, all APs were assumed to be connected to a single central processing unit (CPU), which is responsible to coordinate and process the signals of all UEs in \cite{Ngo2017b,Nayebi2017a,Zhang2018a,Bjornson2019c} (and references therein). That implies that the computational complexity and fronthaul capacity, required for each AP to process and share the data signals related to \emph{all} UEs, grow linearly (or faster) with the number of UEs. Hence, the original form of Cell-Free Massive MIMO was unscalable.
The user-centric approach was reintroduced in \cite{Buzzi2017a,Ngo2018a,Bursalioglu2019a}, but without making the connection to DCC, without analyzing how to utilize it to achieve a provably scalable network operation, and without discussing how it can benefit from the implementation aspects previously addressed in the Network MIMO literature \cite{Bjornson2013d}. This is all done in this paper.

The scalability of the power control algorithms have been considered in a series of previous papers. Particularly, \cite{Ngo2017b,Nayebi2016a,Nayebi2017a,Zhang2018a,Bashar2019a} developed network-wide optimization algorithms for Cell-Free Massive MIMO with a complexity that grows polynomially with the number of APs and UEs. Hence, these algorithms are not feasible for practical implementation; only suboptimal algorithms combined with DCC can be used for power control to achieve a scalable implementation. 
One scalable power control algorithm is ``equal power allocation'', but there are plenty of other/better heuristic algorithms in the literature on both Network MIMO
\cite{Gesbert2007b,Bjornson2010c,Bjornson2011a}, \cite[Sec.~3.4.4]{Bjornson2013d} and Cell-Free Massive MIMO \cite{Buzzi2017a,Nayebi2017a,Interdonato2019a,Nikbakht2019a}.
For a given simulation setup and network utility function, some algorithms will perform better than others, but a numerical evaluation of power control schemes is not the focus of this paper. However, the framework described in this paper, combined with any of the known heuristic power control algorithms, leads to what we call a \emph{Scalable Cell-Free Massive MIMO system}.

\subsection{Motivation and Contributions}
The main motivation of this paper is to define and design scalable Cell-Free Massive MIMO systems. The first step in this direction is taken by proving that Cell-Free Massive MIMO is a special case of the DCC framework from the Network MIMO literature \cite{Bjornson2011a,Bjornson2013d}. This result is instrumental to develop a new scalable algorithm for joint initial access, pilot assignment, and cooperation cluster formation. We are then deriving novel general SE expressions for uplink (UL) and downlink (DL) transmissions of two different cell-free implementations, characterized by different degrees of cooperation among the APs. Inspired by \cite{Bjornson2019c}, the first implementation is a centralized network in which the pilot signals received at all APs are gathered at CPUs, which perform channel estimation, and jointly process the UL and DL data signals. The second implementation is a decentralized network in which each AP locally estimates the channels of its associated UEs and uses this information to locally process data signals. Only the decoding and encoding of data signals is carried out at the CPUs \cite{Bjornson2019c}. For both implementations, the new SE expressions are valid for arbitrary clusters, spatially correlated Rayleigh fading channels, imperfect CSI, APs with any number $N$ of antennas, and heuristic or optimized signal processing schemes. We discuss under which conditions these methods are scalable. A new UL and DL duality is proved and used to heuristically design the DL precoding vectors on the basis of the UL combining vectors.
Numerical results are used to demonstrate that the proposed scalable framework achieves almost the same SE as the state-of-the-art unscalable solutions, and greatly outperforms the original maximum ratio (MR) based Cell-Free Massive MIMO algorithms.

\subsection{Paper Outline and Notation}
The rest of this paper is organized as follows. Section~\ref{sec:network_model} introduces the network model for Cell-Free Massive MIMO and discusses the channel estimation process. Section~\ref{sec:original_cell_free} reviews the system model for the original form of Cell-Free Massive MIMO and then discusses its scalability issues when the number of UEs increases. The DCC framework is reviewed in Section~\ref{sec:DCC_framework}. This section also shows that several cell-free network setups recently proposed can be described by the DCC framework. Section~\ref{sec:scalable_CFmMIMO} proposes a scalable implementation of Cell-Free Massive MIMO for both UL and DL and with two different levels of cooperation among the APs. The performance of the proposed implementations is numerically evaluated and compared in Section~\ref{sec:numerical}. Finally, the main conclusions and implications are drawn in Section~\ref{sec:conclusions}. 

\textit{\textbf{Reproducible research:}} All the simulation results can be reproduced using the Matlab code and data files available at:\\ \url{https://github.com/emilbjornson/scalable-cell-free}

\textit{\textbf{Notation:}} Boldface lowercase letters, $\vect{x}$, denote column vectors and boldface uppercase letters, $\vect{X}$, denote matrices. 
The superscripts $^{\Ttran}$, $^*$, $^{\Htran}$, and $^{\dagger}$ denote transpose, conjugate, conjugate transpose and pseudo-inverse, respectively. 
 The $n \times n$ identity matrix is $\vect{I}_n$.
We use $\triangleq$ for definitions and $\diag(\vect{A}_1,\ldots,\vect{A}_n)$ for a block-diagonal matrix with the square matrices $\vect{A}_1,\ldots,\vect{A}_n$ on the diagonal.
The multi-variate circularly symmetric complex Gaussian distribution with correlation matrix $\vect{R}$ is denoted $\CN(\vect{0},\vect{R})$. The expected value of $\vect{x}$ is denoted as $\mathbb{E}\{ \vect{x} \}$. We use $|\mathcal {A}|$ to denote the cardinality of the set $\mathcal A$.

\section{Network Model}\label{sec:network_model}

\begin{figure}[t!]
	\centering 
	\begin{overpic}[width=\columnwidth,tics=10]{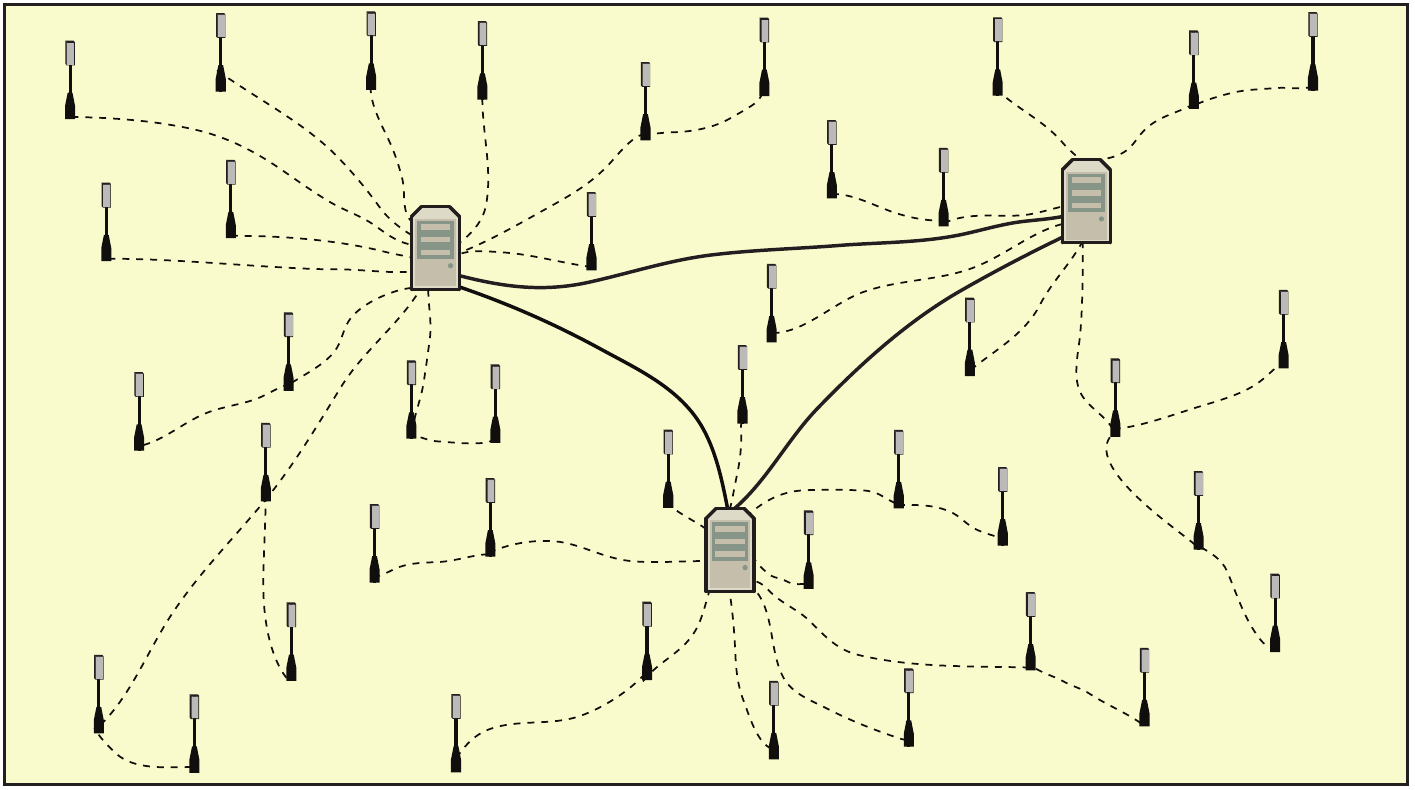}
	\put(80,40){CPU}
	\put(35.5,51){AP $l$}
\end{overpic} 
	\caption{Illustration of a Cell-Free Massive MIMO network with many distributed APs connected to CPUs. The APs may jointly serve the UEs in the coverage area.} \vspace{-2mm}
	\label{fig:cell-free} 
\end{figure}

We consider a cell-free network consisting of $K$ single-antenna UEs and $L$ APs, each equipped with $N$ antennas, that are arbitrarily distributed over the coverage area. The APs are connected to edge-cloud processors, called CPUs \cite{Perlman2015a,Burr2018a,Interdonato2019a}, in an arbitrary fashion. This is illustrated in Fig.~\ref{fig:cell-free} and further discussed in Section~\ref{subsec:network_topology}. This setup enables coherent joint transmission and reception to the UEs in the entire coverage area, and is called Cell-Free Massive MIMO when $L$ and $K$ are large; it is often assumed that $L\gg K$ \cite{Ngo2017b,Nayebi2017a}, but the methods developed in this paper hold for any values.

We assume the APs and UEs operate according to a TDD
protocol with a pilot phase for channel estimation and a data transmission phase. We consider the standard Massive MIMO TDD protocol from \cite{massivemimobook}, where each coherence block is divided into $\tau_p$ channel uses for UL pilots, $\tau_u$ for UL data, and $\tau_d$ for DL data such that $\tau_c = \tau_p + \tau_u + \tau_d$.
The channel between AP $l$ and UE $k$ is denoted by $\vect{h}_{kl} \in \mathbb{C}^N$ and the collective channel from all APs is $\vect{h}_k = [\vect{h}_{k1}^{\Ttran} \, \ldots \, \vect{h}_{kL}^{\Ttran}]^{\Ttran} \in \mathbb{C}^{M}$ with $M = NL$ being the total number of antennas in the coverage area. 
 In each coherence block, an independent correlated Rayleigh fading realization is drawn $\vect{h}_{kl} \sim \CN(\vect{0}, \vect{R}_{kl})$
where $\vect{R}_{kl} \in \mathbb{C}^{N \times N}$ is the spatial  correlation matrix. The Gaussian distribution models the small-scale fading whereas the positive semi-definite correlation matrix $\vect{R}_{kl}$ describes the large-scale fading, including geometric pathloss, shadowing, antenna gains, and spatial channel correlation \cite[Sec. 2.2]{massivemimobook}. We assume the channel vectors of different APs are independently distributed, thus ${\mathbb{E}}\{\vect{h}_{kn}{(\vect{h}_{kl})}^{\Htran}\} = {\bf 0}$ for $l\ne n$. This is a reasonable assumption since the APs are spatially distributed in the network. 
The collective channel is thus distributed as follows:
\begin{equation}\label{eq:collective_channel}
\vect{h}_k \sim \CN(\vect{0}, \vect{R}_{k}) 
\end{equation}
where $\vect{R}_{k} = \diag(\vect{R}_{k1}, \ldots,  \vect{R}_{kL}) \in \mathbb{C}^{M \times M}$
is the block-diagonal spatial correlation matrix. The UEs' channels are independently distributed. We assume the spatial correlation matrices $\{\vect{R}_{kl}\}$ are available wherever needed; see \cite{BSD16A,NeumannJU17,UpadhyaJU17,CaireC17a} for practical correlation matrix estimation methods.

\subsection{Pilot Transmission and Channel Estimation}

We assume there are $\tau_p$ mutually orthogonal  $\tau_p$-length pilot signals, with $\tau_p$ being a constant independent of $K$.\footnote{The value can change over the day, depending on the traffic load, but there will be a maximum value that is supported.} The pilots are assigned to the UEs when they gain access into the network; an algorithm for pilot assignment is proposed in Section~\ref{eq:pilot-assignment}. For now, we let $\mathcal{S}_t \subset \{ 1, \ldots, K\}$ denote the subset of UEs assigned to pilot $t$. When these UEs transmit such a pilot, the received signal $\vect{y}_{tl}^{\rm{pilot}} \in \mathbb{C}^N$ after despreading at AP $l$ is \cite[Sec. 3]{massivemimobook}
\begin{equation}\label{eq:received_uplink_pilot}
\vect{y}_{tl}^{\rm{pilot}} = \sum_{i \in \mathcal{S}_t} \sqrt{ \tau_p p_i} \vect{h}_{il} + \vect{n}_{tl}
\end{equation}
where $p_i$ is the transmit power of UE $i$, $\tau_p$ is the processing gain, and $\vect{n}_{tl} \sim \CN (\vect{0}, \sigma^2 \vect{I}_N)$ is the thermal noise. Using standard results \cite[Sec.~3]{massivemimobook}, the minimum mean-squared-error (MMSE) estimate of $\vect{h}_{kl}$ for $k \in \mathcal{S}_t$ is  
\begin{equation} \label{eq:estimates}
\widehat{\vect{h}}_{kl} = \sqrt{p_k \tau_p} \vect{R}_{kl} \vect{\Psi}_{tl}^{-1} \vect{y}_{tl}^{\rm{pilot}}
\end{equation}
where \begin{equation} \label{eq:Psitl}
\vect{\Psi}_{tl} = \mathbb{E} \{ \vect{y}_{tl}^{\rm{pilot}} (\vect{y}_{tl}^{\rm{pilot}})^{\Htran} \} = \sum_{i \in \mathcal{S}_t} \tau_p p_i \vect{R}_{il} + \sigma^2 \vect{I}_{N}
\end{equation}
is the correlation matrix of \eqref{eq:received_uplink_pilot}.
The mutual interference generated in \eqref{eq:received_uplink_pilot} by the pilot-sharing UEs in $\mathcal{S}_t $ causes the so-called \emph{pilot contamination} that degrades the system performance, similar to the case in standard Massive MIMO. Pilot contamination has two main consequences \cite[Sec.~3.3.2]{massivemimobook}: firstly, it reduces the estimation quality that makes coherent transmission less effective; secondly, the estimates $\widehat{\vect{h}}_{kl}$ for $k \in \mathcal{S}_t$ become correlated, which leads to additional interference.
Both effects have an impact on the UEs' performance but it is only the second one that is responsible of the so-called \emph{coherent interference} \cite[Sec.~4.2]{massivemimobook}, which has received particular attention in the literature since it might increase linearly with $N$, just as the signal term \cite{Marzetta2010a,Yin2016a,BjornsonHS17}.

Notice that the $N\times N$ matrix $\sqrt{p_k \tau_p} \vect{R}_{kl} \vect{\Psi}_{tl}^{-1}$ in \eqref{eq:estimates} only depends on the channel statistics, which by definition are fixed throughout the communication. The matrix can thus be precomputed at AP $l$ with negligible complexity.\footnote{In practice, the statistics change due to UE mobility or new scheduling decisions, but this is outside the scope of this paper.} The
MMSE estimation requires to first compute $\vect{y}_{tl}^{\rm{pilot}}$ and then multiply it with the precomputed statistical matrix $\sqrt{p_k \tau_p} \vect{R}_{kl} \vect{\Psi}_{tl}^{-1}$ of each UE served by AP $l$. The first operation requires $N\tau_p$ complex multiplications per pilot sequence while the second needs $N^{2}$ complex multiplications per UE \cite[Sec.~3.4]{massivemimobook}.

\section{Definition of Scalability and Review of the Original Form of Cell-Free Massive MIMO}\label{sec:original_cell_free}

To motivate and better understand the scalable framework proposed in this paper, we first review the system model of the original Cell-Free Massive MIMO \cite{Ngo2017b,Nayebi2017a,Nayebi2016a}, where there are many APs that all of them simultaneously serve all the UEs in the system in both UL and DL. 

\subsection{Uplink and Downlink Data Transmissions}

During UL data transmission, the received signal ${\bf y}_{l}^{\rm{ul}} \in \mathbb {C}^{N}$ at AP $l$ is
\begin{align}
	{\bf y}_{l}^{\rm{ul}} = \sum\limits_{i=1}^{K} {\bf h}_{il}s_{i} + {\bf n}_{l}\label{y_{l}^{ul}}
\end{align}
where $s_{i} \in \mathbb{C}$ is the signal transmitted from UE $i$ with power $p_i$ and ${\bf n}_{l}  \sim \mathcal {CN}({\bf 0},\sigma^2{\bf I}_{N})$. Network-wide UL decoding was considered in the original papers on Cell-Free Massive MIMO  \cite{Ngo2017b,Nayebi2016a}. In that case, AP $l$ selects a receive combining vector $\vect{v}_{kl} \in \mathbb {C}^{N}$ for UE $k$ and computes $\vect{v}_{kl}^{\Htran}{\bf y}_{l}^{\rm{ul}}$ locally. The network then estimates $s_k$ by computing the summation
\begin{align}\notag
\!\!\widehat{s}_k &= \sum_{l=1}^{L}  
\vect{v}_{kl}^{\Htran}{\bf y}_{l}^{\rm{ul}} \\ \notag & = \left(\sum_{l=1}^{L} 
\vect{v}_{kl}^{\Htran}{\bf h}_{kl}\right)s_{k}  + \!\!\!\sum\limits_{i=1,i\ne k}^{K} \!\!\left(\sum_{l=1}^{L} 
 \vect{v}_{kl}^{\Htran}{\bf h}_{il}\right)s_{i} + \sum_{l=1}^{L} 
\vect{v}_{kl}^{\Htran}{\bf n}_{l}\\
&= 
\vect{v}_{k}^{\Htran}{\bf h}_{k}s_{k}  + \!\!\!\sum\limits_{i=1,i\ne k}^{K} 
\!\!\! \vect{v}_{k}^{\Htran}{\bf h}_{i}s_{i} + \vect{v}_{k}^{\Htran}{\bf n}
\label{eq:Cell-free_UL}
\end{align}
where $\vect{v}_k = [\vect{v}_{k1}^{\Ttran} \, \ldots \, \vect{v}_{kL}^{\Ttran}]^{\Ttran} \in \mathbb{C}^{M}$ denotes the collective combining vector and $\vect{n} = [\vect{n}_{1}^{\Ttran} \, \ldots \, \vect{n}_{L}^{\Ttran}]^{\Ttran} \in \mathbb{C}^{M}$ collects all the noise vectors.
Notice that \eqref{eq:Cell-free_UL} is equivalent to a UL single-cell Massive MIMO system model with combined channels $\{\vect{v}_{k}^{\Htran}{\bf h}_{i}= \sum_{l=1}^{L} 
 \vect{v}_{kl}^{\Htran}{\bf h}_{il}: i=1,\ldots,K\}$. Therefore, the achievable UL SEs easily follow from that literature \cite[Sec. 4.1]{massivemimobook}. The key difference between cell-free and cellular networks lies in the design of combining vectors $\{{\bf v}_{kl}\}$ since each AP should preferably only use CSI that it can acquire locally in the pilot transmission phase, which is referred to as local CSI. The most popular choice in the Cell-Free Massive MIMO literature is MR combining with $ \vect{v}_{kl} = \widehat{\vect{h}}_{kl}$ \cite{Ngo2017b,Buzzi2017a,Zhang2018a}. Other heuristic combining schemes such as local MMSE (L-MMSE) have also been considered \cite{Bjornson2019c}. Network-wide UL power optimization methods can be found in \cite{Ngo2017b,Nayebi2016a,Bashar2019a}, among others.

Let $\vect{w}_{il} \in \mathbb{C}^N$ denote the precoder that AP $l$ assigns to UE $i$. During DL transmission, the received signal at UE $k$ is
\begin{align}
y_k^{\rm{dl}} = \sum_{l=1}^{L} \vect{h}_{kl}^{\Htran} \sum_{i=1}^{K}  \vect{w}_{il} \varsigma_i + n_k  
=  \vect{h}_k^{\Htran}\sum_{i=1}^{K}  \vect{w}_i \varsigma_i + n_k \label{eq:Cell-free}
\end{align}
where $\varsigma_i \in \mathbb{C}$ is the independent unit-power data signal intended for UE $i$ (i.e., $\mathbb{E} \{{\|\varsigma_i \|^2}\} = 1$), $\vect{w}_k = [\vect{w}_{k1}^{\Ttran} \, \ldots \, \vect{w}_{kL}^{\Ttran}]^{\Ttran} \in \mathbb{C}^{M}$ is the collective precoding vector, and $n_k \sim \CN(0,\sigma^2)$ is the receiver noise. Since $ \vect{h}_k$ is distributed as in \eqref{eq:collective_channel}, the system model \eqref{eq:Cell-free} is mathematically equivalent to a DL single-cell Massive MIMO system with correlated fading \cite[Sec. 2.3.1]{massivemimobook}. Therefore, the achievable DL SE in Cell-Free Massive MIMO follows easily from the literature on Massive MIMO with correlated fading \cite[Sec. 4.3]{massivemimobook}. As in the UL, the main difference between cell-free and cellular networks is in the selection of the precoding vectors, which 
should use only local CSI and also satisfy per-AP power constraints. The most popular choice is MR precoding with 
\begin{align}\label{eq:MR-precoding}
\vect{w}_{il} = \sqrt{\rho_{i}} \, \frac{\widehat{\vect{h}}_{il}}{ \sqrt{\mathbb{E} \{{\|\widehat{\vect{h}}_{il}\|^2}\}}}
\end{align}
where $\rho_{i}\geq 0$ is the transmit power allocated to UE $i$. It can also be adapted using the network-wide DL power optimization methods developed (among others) in \cite{Ngo2017b,Nayebi2017a,Zhang2018a}.

\subsection{The Scalability Issue}

Although the network-wide processing in the original Cell-Free Massive MIMO is appealing, it is not practical for large-scale network deployments with many UEs. To determine if a network technology is scalable or not, it is helpful to let $K \to \infty$ and see which of the following tasks remain practically implementable:
\begin{enumerate}
\item \emph{Signal processing for channel estimation};
\item\emph{Signal processing for data reception and transmission};
\item \emph{Fronthaul signaling for data and CSI sharing};
\item \emph{Power control optimization}.
\end{enumerate}

Based on this list, we make the following definition.

\begin{definition}[Scalability] \label{def:scalable}
A Cell-Free Massive MIMO network is \emph{scalable} if all the four above-listed tasks have finite complexity and resource requirements for each AP as $K \to \infty$.
\end{definition}

The original form of Cell-Free Massive MIMO fails to be scalable with respect to all of the four above-listed tasks:
\begin{enumerate}
\item AP $l$ must compute channel estimates $\{\widehat{\vect{h}}_{kl}: k=1,\ldots,K\}$ for all $K$ UEs, which has an infinite complexity as $K \to \infty$.
\item AP $l$ needs to create the transmitted signal $\sum_{k=1}^{K} \vect{w}_{kl} \varsigma_i$ where the summation implies infinite complexity as $K \to \infty$. The complexity of computing the $K$ precoding vectors $\{\vect{w}_{kl}: k=1,\ldots,K\}$ depends on the precoding scheme, but even the low-complexity MR precoding scheme requires that the AP takes the $K$ channel estimates and normalizes each one as in \eqref{eq:MR-precoding} to satisfy the power constraint.
The same scalability issue appears in the UL, where the AP needs to compute $\{\vect{v}_{kl}^{\Htran}{\bf h}_{kl}: k=1,\ldots,K\}$ using $K$ different combining vectors $\{\vect{v}_{kl}: k=1,\ldots,K\}$.

\item  AP $l$ needs to receive the $K$ DL data signals $\{\varsigma_k: k=1,\ldots,K\}$ from a CPU and must forward its $K$ processed received signals $\{\vect{v}_{kl}^{\Htran}{\bf y}_{l}^{\rm{ul}}: k=1,\ldots,K\}$ over the fronthaul links. The number of scalars to be sent over the fronthaul grows unboundedly as $K \to \infty$.
\item Any non-trivial network-wide power optimization has a complexity that goes to infinity as $K \to \infty$. For example, the complexity of solving linear or convex optimization problems grows polynomially in the number of optimization variables.
\end{enumerate}
\smallskip

In the remainder of this paper, we develop a novel implementation framework that is scalable according to Definition~\ref{def:scalable}. Although we start from the DCC framework for Network MIMO that was proposed in \cite{Bjornson2011a} and claimed to be scalable by the authors, we need to fill in many missing details to make it truly scalable (e.g., dealing with initial access, pilot assignment, and channel estimation).

\begin{remark}
The considered scalability definition keeps the computational complexity and fronthaul resources per AP finite as $K \to \infty$. The total computational complexity and fronthaul requirements will then be independent of $K$ but proportional to $L$. Since $LN \gg K$ for the network to be practically useful, $K \to \infty$ requires also $L \to \infty$. This means the total complexity/requirements will diverge but, as long as each AP is equipped with a local processor and a fronthaul connection of sufficient, but finite, capacity, each AP can carry out its necessary tasks irrespective of how large the network is. Hence, the scalability definition is practically sound.
\end{remark}

\section{Dynamic cooperation clustering}\label{sec:DCC_framework}

The DCC framework was proposed in \cite{Bjornson2011a,Bjornson2013d} to enable ``\emph{unified analysis of anything from interference channels to Network MIMO}''. This is achieved by defining a set of diagonal matrices $\vect{D}_{il} \in \mathbb{C}^{N \times N}$, for $i=1,\ldots,K$ and $l=1,\ldots,L$, determining which AP antennas may transmit to which UEs. More precisely, the $j$th diagonal element of $\vect{D}_{il}$ is 1 if the $j$th antenna of AP $l$ is allowed to transmit to and decode signals from UE $i$ and 0 otherwise. In this section, we will show that the original Cell-Free Massive MIMO in \eqref{eq:Cell-free_UL} and \eqref{eq:Cell-free} is one of the many setups that can be described by the DCC framework.

\begin{figure}[t!]
	\centering 
	\begin{overpic}[width=\columnwidth,tics=10]{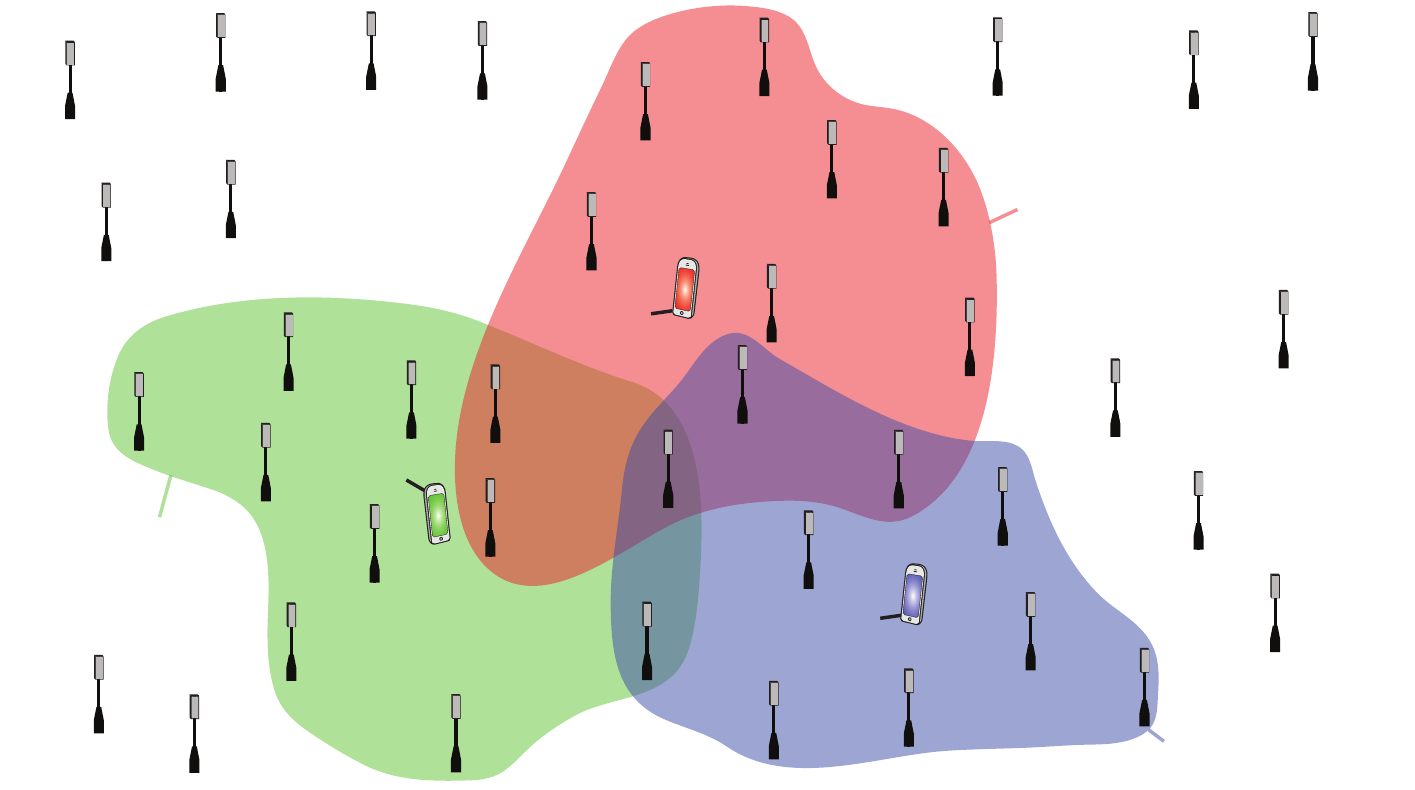}
	\put(38,32){\footnotesize{UE 1}}
	\put(73,41){\footnotesize{AP cluster}}
	\put(73,38){\footnotesize{for UE 1}}
	\put(21,22){\footnotesize{UE 2}}
	\put(1,16){\footnotesize{AP cluster}}
	\put(1,13){\footnotesize{for UE 2}}
	\put(54.5,10){\footnotesize{UE 3}}
	\put(83,5){\footnotesize{AP cluster}}
	\put(83,2){\footnotesize{for UE 3}}
\end{overpic}  \vspace{-3mm}
	\caption{Example of dynamic cooperation clusters for three UEs in a cell-free network with a large number of APs.}
	\label{fig:illustrateCooperation} 
\end{figure}
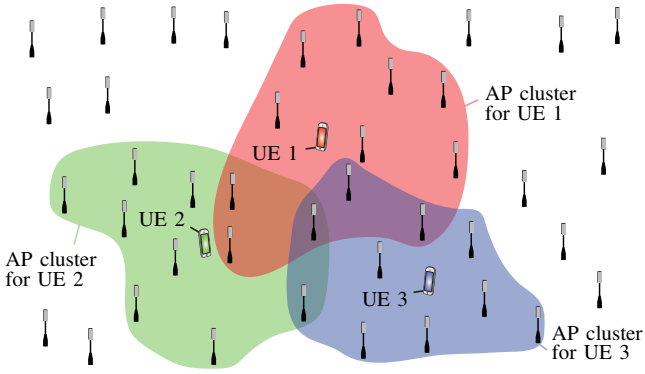

Fig.~\ref{fig:illustrateCooperation} illustrates a network with three UEs that are served by a large number of APs. The colored regions illustrate which clusters of APs are serving which UEs and (implicitly) determine the matrices  $\vect{D}_{il}$. The fact that the clusters are partially overlapping is the core feature of DCC and also demonstrates that it is a cell-free network.

\subsection{Uplink and Downlink Data Transmissions}
The DCC framework does not change the received UL signal ${\bf y}_{l}^{\rm{ul}}$ in $\eqref{y_{l}^{ul}}$ since all APs will physically receive the signal from all UEs. However, only a subset of the APs are taking part in the signal detection, thus \eqref{eq:Cell-free_UL} changes to 
\begin{align}  \notag
\widehat{s}_k &= \sum_{l=1}^{L} 
\vect{v}_{kl}^{\Htran} \vect{D}_{kl} {\bf y}_{l}^{\rm{ul}} \\ &= 
\vect{v}_{k}^{\Htran}\vect{D}_{k}{\bf h}_{k}s_{k}  + \!\!\!\sum\limits_{i=1,i\ne k}^{K} 
\!\!\! \vect{v}_{k}^{\Htran}\vect{D}_{k}{\bf h}_{i}s_{i} + \vect{v}_{k}^{\Htran}\vect{D}_{k}{\bf n}\label{eq:uplink-data-estimate}
\end{align}
where $\vect{D}_k = \diag(\vect{D}_{k1}, \ldots, \vect{D}_{kL}) \in \mathbb{C}^{M \times M}$ is a block-diagonal matrix.

The received DL signal at UE $k$ in \eqref{eq:Cell-free} becomes
\begin{align} 
y_k^{\rm{dl}} = \sum_{l=1}^{L}  \vect{h}_{kl}^{\Htran}\sum_{i=1}^{K} \vect{D}_{il}  \vect{w}_{il} \varsigma_i + n_k = \vect{h}_k^{\Htran} \sum_{i=1}^{K} \vect{D}_i \vect{w}_i \varsigma_i + n_k.\label{eq:DCC}
\end{align}
Note that $ \vect{D}_{il}=\vect{0}_N$ implies $\vect{D}_{il}  \vect{w}_{il}=\vect{0}_N$ and $\vect{v}_{il}^{\Htran} \vect{D}_{il} =\vect{0}_N$, thus only APs with $ \vect{D}_{il} \neq \vect{0}_N$ will transmit to UE $k$ in the DL and apply receive combining in the UL.
By selecting $\{\vect{D}_1,\ldots,\vect{D}_K\}$ in different ways,  \eqref{eq:uplink-data-estimate} and \eqref{eq:DCC} can be used to model many different types of networks with multiple APs; see \cite[Sec.~1]{Bjornson2013d} for examples that go beyond Cell-Free Massive MIMO.

The original Cell-Free Massive MIMO in \eqref{eq:Cell-free_UL} and \eqref{eq:Cell-free} is obtained from \eqref{eq:uplink-data-estimate} and \eqref{eq:DCC}, respectively, in the special case of $\vect{D}_i=\vect{I}_M \, \forall i$, where all AP antennas serve all UEs.
The user-centric approach to Cell-Free Massive MIMO described in \cite{Buzzi2017a} is also an instance of the DCC framework. In \cite{Buzzi2017a}, $\mathcal{M}_i \subset \{ 1, \ldots, L\}$ denotes the subset of APs that serve UE $i$ and we adopt the same notation in this paper. This corresponds to setting
\begin{equation}\label{eq:user-centric}
\vect{D}_{il}= \begin{cases}
\vect{I}_N & \textrm{if } l \in \mathcal{M}_i \\ 
\vect{0}_N & \textrm{if } l \not \in \mathcal{M}_i \end{cases}
 \end{equation}
which is exactly the setup previously considered in \cite{Bjornson2011a}.
The Fog Massive MIMO architecture, described for DL-only data transmission in \cite{Bursalioglu2019a}, is also an instance of the DCC framework; the only difference is that $\mathcal{A}_i$, instead of $\mathcal{M}_i$, is used to denote the subset of APs that serve UE $i$.

\subsection{A Partial Solution to the Scalability Issue}

The DCC framework was proposed in \cite{Bjornson2011a} to achieve scalability in Network MIMO, but without proving this claim mathematically or taking imperfect CSI into account.
In this paper, we provide these important missing details with particular focus on Cell-Free Massive MIMO. To this end, we first define the set of UEs served by at least one of the antennas at AP $l$:
\begin{equation}
\mathcal{D}_l = \bigg\{ i : \tr(\vect{D}_{il})\geq 1, i \in \{ 1, \ldots, K \} \bigg\}.
\end{equation}
According to Definition~\ref{def:scalable}, a sufficient condition for scalability is as follows.

\begin{lemma} \label{lemma:D-requirement}
If the cardinality $|\mathcal{D}_l |$ is constant as $K \to \infty$ for $l=1,\ldots,L$, then the Cell-Free Massive MIMO network with DCC satisfies the first three conditions in Definition~\ref{def:scalable}.
\end{lemma}
\begin{IEEEproof}
AP $l$ only needs to compute the channel estimates and precoding/combining vectors for $|\mathcal{D}_l |$ UEs. This has a constant complexity as $K \to \infty$ if $|\mathcal{D}_l |$ is constant. Moreover, AP $l$ only needs to send/receive data related to these $|\mathcal{D}_l |$ UEs over the fronthaul network, which is a constant number as $K \to \infty$.
\end{IEEEproof}

The only part of Definition~\ref{def:scalable} that is not captured by Lemma~\ref{lemma:D-requirement} is the complexity of the power optimization, but this fourth condition is relatively easy to satisfy; for example, by using full power in UL and equal power allocation between the served UEs in DL. With these results in mind, the practically important question is how to select the sets $\mathcal{D}_l $ for $l=1,\ldots,L$ in a scalable way, while guaranteeing service to all UEs. This challenge is tackled in the next section, along with the design of combining and precoding schemes for a scalable Cell-Free Massive MIMO system.

\section{Scalable Cell-Free Massive MIMO Framework}\label{sec:scalable_CFmMIMO}
We will now propose the first scalable implementation of Cell-Free Massive MIMO. It is inspired by the guidelines for distributed Network MIMO in \cite[Sec.~4.3, 4.7]{Bjornson2013d}, but is substantially more detailed and also focused on channel estimation and resource allocation, particularly pilot allocation.
We first note that the algorithms for user-centric clustering in previous works have two deficiencies: \cite{Ngo2018a,Interdonato2019a} do not limit how many UEs that an AP can serve, making them unscalable as $K \to \infty$, and \cite{Buzzi2017a,Bursalioglu2019a} do not guarantee that all UEs are served.
We will develop an algorithm for joint initial access, pilot assignment, and cluster formation that resolves both issues. We first make the following key assumption.

\begin{assumption} \label{assumption1}
Each AP serves at most one UE per pilot sequence and uses all its $N$ antennas to serve these UEs. 
\end{assumption}

The above assumption implies that $|\mathcal{D}_l |  \leq \tau_p$ and 
\begin{equation}\label{eq:def_D_il}
\vect{D}_{il}= \begin{cases}
\vect{I}_N & i \in \mathcal{D}_l \\
\vect{0}_N & i \not \in \mathcal{D}_l
\end{cases}
\end{equation}
for  $l=1,\ldots,L$. Since $\tau_p$ was assumed to be independent of $K$, the scalability requirement in Lemma~\ref{lemma:D-requirement} is satisfied.
The rationale behind Assumption~\ref{assumption1} is as follows:
\begin{enumerate}
\item Pilot contamination degrades the channel estimation quality and causes coherent interference \cite{massivemimobook}. If an AP serves more than one UE per pilot sequence, the signals from and to these pilot-sharing UEs will be strongly interfering, which is undesired.
\item The channel estimation and signal processing (i.e., precoding and combining) complexity become fixed and scalable, even if all $N$ antennas are used.
\item The fronthaul links need to support $\tau_p$ parallel UL and DL data signals per AP.
\end{enumerate}

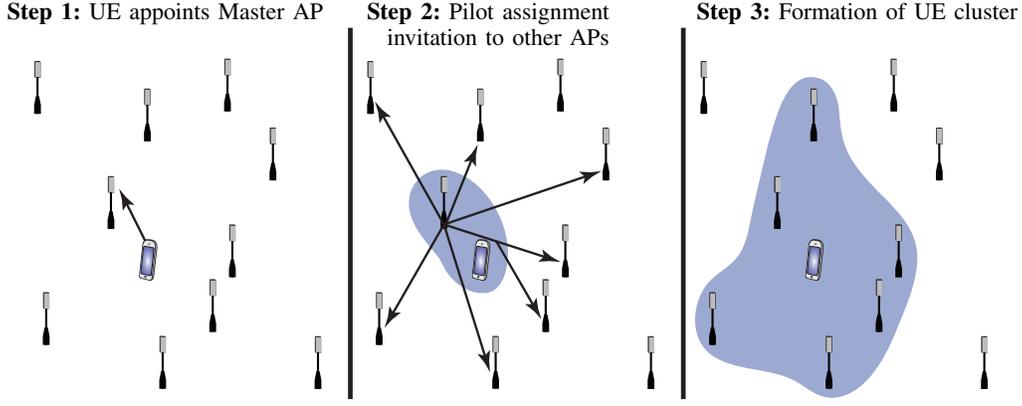
\begin{figure*}[t!]
	\centering 
	\begin{overpic}[width=1.5\columnwidth,tics=10]{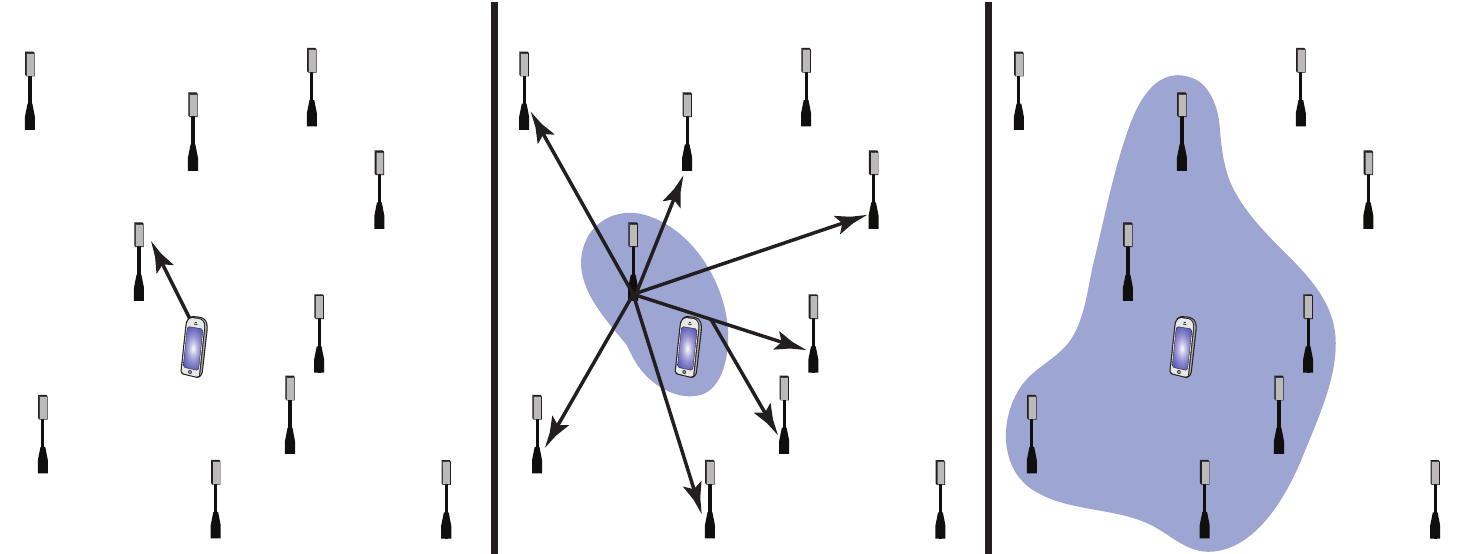}
	\put(-1,38){\small{\textbf{Step 1:} UE appoints Master AP}}
	\put(35,38){\small{\textbf{Step 2:} Pilot assignment}}
	\put(37,35.5){\small{invitation to other APs}}
	\put(68,38){\small{\textbf{Step 3:} Formation of UE cluster}}
	\end{overpic} 
	\caption{The proposed procedure for joint initial access, pilot assignment, and cluster formation consists of three main steps.}
	\label{fig:cellfree_formation}  
\end{figure*}

\subsection{Algorithm for Joint Initial Access, Pilot Assignment, and Cluster Formation} \label{eq:pilot-assignment}

When a new UE wants to access the network, it needs to be assigned a pilot and make it into the set $\mathcal{D}_l $ of at least one AP. This must be done in a distributed fashion, which has the risk that the UE is inadvertently dropped from service since no AP decides to transmit to it \cite{Buzzi2017a,Bursalioglu2019a}. To avoid that, we propose the three-step access procedure illustrated in Fig.~\ref{fig:cellfree_formation}. In particular, each UE appoints a \emph{Master AP} that is required to transmit to it in the DL and coordinate the decoding of the UL data. If $K+1$ is the index of the connecting UE, then the proposed algorithm is:
\smallskip
\begin{enumerate}

\item[Step 1:] The accessing UE measures $\beta_{l} = \tr(\vect{R}_{(K+1)l})/N$ for all nearby APs by using the periodically broadcasted synchronization signals.\footnote{Such signals exist in standards for cellular networks. Examples are the primary and secondary synchronization signals in 5G.} Then, the UE appoints AP $\ell$ with
\begin{align}
\ell = \mathrm{arg \, max}_{l} \, \beta_{l} 
\end{align}
as its Master AP. The UE also uses the broadcasted signal to synchronize to the AP. The UE contacts its Master AP via a standard random access procedure \cite{Sesia09,Sanguinetti12,sanguinetti13bis}. 

\item[Step 2:]  The appointed Master AP responds by assigning pilot  $\tau$ to the UE, where
\begin{align} \label{eq:best-pilot}
\tau = \mathrm{arg \, min}_{t} \, \tr( \vect{\Psi}_{tl} )
\end{align}
with $\vect{\Psi}_{tl}$ given in \eqref{eq:Psitl}. The computation only involves the existing $K$ UEs and $\tau$ is the pilot where the Master AP observes the least pilot contamination. The Master AP informs a limited set of neighboring APs that it is going to serve UE $K+1$ on pilot $\tau$.

\item[Step 3:] Each of the neighboring AP decides if it will serve UE $K+1$. The decision rule is to serve the UE if either the AP does not serve any UE on pilot $\tau$ or the new UE has a better channel than the UE it currently serves, in which case it switches serving to the new UE.

\end{enumerate}

\smallskip
In summary, the UE appoints the AP with the strongest large-scale fading channel coefficient as its Master AP and it is assigned to the pilot that this AP observes the least pilot power on.\footnote{This should be a pilot on which the AP is not currently serving a UE as being its Master AP, since that role has higher priority. Each AP can only be the Master AP of up to $\tau_p$ UEs in the proposed framework, but in the unlikely event that this cannot be satisfied, multiple UEs can be assigned to the same pilot but multiplexed in time and/or frequency instead.}

When a UE moves around or other UEs leave or connect to the network, the proposed access procedure can be redone. The UE can initiate such a procedure by appointing a new Master AP, in which case the procedure proceeds as if the UE would be inactive and is now accessing the network. The new Master AP then informs the previous Master AP that it has taken over the service of the UE. Alternatively, the current Master AP can periodically update the pilot assignment by computing \eqref{eq:best-pilot} and check if this pilot would lead to less pilot contamination than the pilot currently assigned to the UE. In that case, the pilot is changed.

The AP uses the MMSE channel estimates for receiving the UL data and for precoding the DL data. AP $l$ only needs to compute estimates $\widehat{\vect{h}}_{kl}$ for $k \in \mathcal{D}_l $, which under Assumption~\ref{assumption1} is at most one UE per pilot. Since the complexity per AP is independent of $K$, the initial access and pilot assignment are scalable when $K \to \infty$.

\begin{remark} 
A weakness of the system model in \eqref{eq:DCC} is that all transmissions are perfectly synchronized, which is practically infeasible due to propagation delays. However, if every AP is synchronized with its neighbors and the UE synchronizes with its Master AP in the initial access, the dominant terms in \eqref{eq:DCC} will be accurate, making the model reasonable for analysis.
\end{remark}

\subsection{Uplink Data Transmission}
Next, we derive and analyze achievable UL SE expressions for the DCC signal model in \eqref{eq:uplink-data-estimate}.
We assume the APs delegate the task of UL data decoding to a nearby CPU with high computational resources (see Section~\ref{subsec:network_topology} for details). Two levels of cooperation among the APs are described below and compared in terms of achievable SE and fronthaul signaling load. In the first level, the pilot and data signals received at the APs are gathered (through the fronthaul links) at the CPU, which performs channel estimation and data detection in a fully centralized fashion. In the second level, AP $l$ uses the available estimates $\{\widehat{\vect{h}}_{il} : i \in \mathcal{D}_l \}$ and the UL signal ${\bf y}_{l}^{\rm{ul}}$ in $\eqref{y_{l}^{ul}}$ to compute soft estimates $\vect{v}_{kl}^{\Htran}{\bf y}_{l}^{\rm{ul}}$ of data, which are sent to the CPU for final decoding. 

\subsubsection{Centralized combining}\label{centralized_combining_UL}
The most advanced level of cooperation among APs for decoding the signal from UE $k$ is when AP $l\in\mathcal{M}_k \subset \{ 1, \ldots, L\}$ sends its pilot signals $\{ \vect{y}_{tl}^{\rm{pilot}}: t=1,\ldots,\tau_p\}$ and data signal $\vect{y}_{l}^{\rm{ul}}$ to the CPU, which takes care of channel estimation and data detection in a centralized fashion. In other words, the APs act as relays that forward all signals to the CPU \cite{Estella2019a}. In each coherence block, AP $l$ needs to send $\tau_p N$ complex scalars for the pilot signals and $\tau_uN$ complex scalars for the received signals. This is summarized in Table~\ref{tab:signaling}.

\begin{table}[t]  
        \caption{Number of complex scalars that a generic AP $l$ needs to exchange with the CPU over the fronthaul in each coherence block in the centralized and distributed cases.}  \label{tab:signaling}
\centering
    \begin{tabular}{|c|c|c|c|} 
    \hline
    {} & {Pilot signals}&{Uplink signals}&{Downlink signals}   \\      \hline\hline 

   Centralized&   $\tau_p N$  & $ \tau_u N$  &  $\tau_d N$    \\ \hline    
   Distributed&   ---  & $\tau_u|\mathcal{D}_l| $ & $\tau_d|\mathcal{D}_l| $  \\   \hline    
    \end{tabular}
\end{table}

In the above circumstances, the signal available at the CPU for decoding UE $k$ is $\widehat s_k$ given in \eqref{eq:uplink-data-estimate}. 
The CPU can select an arbitrary receive combining vector $\vect{v}_k \in \mathbb{C}^{LN}$ for UE $k$ based on its CSI. The collective channel estimate $\widehat{\vect{h}}_{i} = [ \widehat{\vect{h}}_{i1}^{\Ttran} \, \ldots \, \widehat{\vect{h}}_{iL}^{\Ttran} ]^{\Ttran}$ can be only partially computed since only some APs send their received pilot signals. More precisely, the CPU knows
\begin{equation} \label{eq:estimates-central}
\vect{D}_i \widehat{\vect{h}}_{i} 
= \begin{bmatrix} \vect{D}_{i1} \widehat{\vect{h}}_{i1} \\ \vdots \\  \vect{D}_{iL} \widehat{\vect{h}}_{iL} 
\end{bmatrix} 
\sim \CN \left( \vect{0}, p_i \tau_p \vect{D}_{i} \vect{R}_{i} \vect{\Psi}_{t_i}^{-1} \vect{R}_{i} \vect{D}_{i} \right)
\end{equation}
where $\vect{\Psi}_{t_i}^{-1} = \diag( \vect{\Psi}_{t_i1}^{-1}, \ldots, \vect{\Psi}_{t_iL}^{-1})$ and unknown estimates are represented by zeros. For later use, we define the collective estimation error $\tilde{\vect{h}}_i = \vect{h}_i - \widehat{\vect{h}}_{i} \sim \CN(\vect{0},\vect{C}_i)$ with
$\vect{C}_i = \diag(\vect{C}_{i1}, \ldots, \vect{C}_{iL})$.

The ergodic capacity is unknown for this setup, but we can rigorously analyze the performance by using standard capacity lower bounds
\cite{Biglieri1998a,massivemimobook}, which we refer to as achievable SEs.

\begin{proposition} \label{theorem:uplink-capacity-general}
When the MMSE channel estimates are known, an achievable SE of UE $k$ is
\begin{equation} \label{eq:uplink-rate-expression-general}
\mathacr{SE}_{k}^{({\rm ul},1)} = \frac{\tau_u}{\tau_c} \mathbb{E} \left\{ \log_2  \left( 1 + \mathacr{SINR}_{k}^{({\rm ul},1)}  \right) \right\}
\end{equation}
where the instantaneous effective signal-to-interference-and-noise ratio (SINR) is given by
\begin{equation} \label{eq:uplink-instant-SINR}
 \mathacr{SINR}_{k}^{({\rm ul},1)} = \frac{ p_{k} |  \vect{v}_{k}^{\Htran} \vect{D}_{k}\widehat{\vect{h}}_{k} |^2  }{ 
\sum\limits_{i=1,i\ne k}^K p_{i} | \vect{v}_{k}^{\Htran} \vect{D}_{k}\widehat{\vect{h}}_{i} |^2
+ \vect{v}_{k}^{\Htran}  {\bf{Z}}_k \vect{v}_{k}  
}
\end{equation}
with ${\bf{Z}}_k =  \vect{D}_{k}\left(\sum\limits_{i=1}^{K} p_{i}  \vect{C}_{i}  + \sigma_{\rm ul}^2  \vect{I}_{LN} \right)\vect{D}_{k}$.
\end{proposition}
\begin{IEEEproof}
It follows the proof of \cite[Th.~4.1]{massivemimobook} for Massive MIMO and is therefore omitted.
\end{IEEEproof}

The SE expression in \eqref{eq:uplink-rate-expression-general} holds for any receive combining vector $\vect{v}_{k}$, spatial correlation matrices,  and selection of the DCC, and is thus a generalization of \cite{Nayebi2016a,Chen2018b,Bashar2018a,Bjornson2019c}.
The expression can be easily computed for any $\vect{v}_{k}$ by using Monte Carlo methods, as done in Section~\ref{sec:numerical}. 
A possible choice is to use MR combining with 
\begin{align}\label{MR_combining_D_k}
\vect{v}_k^{{\rm MR}} = \vect{D}_{k}\widehat{\vect{h}}_k 
\end{align}
which has low computational complexity and maximizes the power of the desired signal in the numerator of \eqref{eq:uplink-instant-SINR}, but the downside is that it neglects the existence of interference in the denominator. MR was considered in the original Cell-Free Massive MIMO papers  \cite{Ngo2017b,Nayebi2016a}. Table~\ref{tab:cost_linear_processing} summarizes the computational complexity\footnote{This is obtained from \cite[Sec. 4.1.2]{massivemimobook} under the assumption that all the statistical matrices have already been precomputed and stored, since they are constant throughout the communication.} with MR combining in terms of number of complex multiplications per UE. In deriving those numbers, we have taken into account that $\vect{D}_k = \diag(\vect{D}_{k1}, \ldots, \vect{D}_{kL})$ in \eqref{MR_combining_D_k} and that $ \vect{D}_{kl}=\vect{0}_N$ implies $\vect{D}_{kl} \widehat{ \vect{h}}_{kl}=\vect{0}_N$. Hence, only the MMSE channel estimates $\widehat{\vect{h}}_{kl}$ with $l\in \mathcal{M}_k$ are needed to compute $\vect{v}_k^{{\rm MR}}$, while all other APs assign a zero-valued receive combining vector.

Despite the low complexity, MR combining is known to be a vastly suboptimal scheme \cite{Nayebi2016a,Bjornson2019c}. To obtain a good and scalable solution, we first notice that $\mathacr{SINR}_{k}^{({\rm ul},1)}$ in \eqref{eq:uplink-instant-SINR} has the form of a generalized Rayleigh quotient. Hence, the optimal combining can be obtained as follows.

\begin{corollary} \label{cor:MMSE-combining}
The instantaneous SINR in \eqref{eq:uplink-instant-SINR} for UE~$k$ is maximized by the MMSE combining vector
\begin{equation} \label{eq:MMSE-combining}
\vect{v}_{k}^{{\rm MMSE}} =    p_{k} \left( \sum\limits_{i=1}^{K} p_{i}  \vect{D}_{k}\widehat{\vect{h}}_{i} \widehat{\vect{h}}_{i}^{\Htran}\vect{D}_{k} + {\bf{Z}}_k \right)^{\!\dagger} \!\!\vect{D}_{k}\widehat{\vect{h}}_{k}
\end{equation}
which leads to the maximum value
\begin{align} \label{eq:uplink-maximized-SINR}
\!\!\!\!\!\!\mathacr{SINR}_{k}^{({\rm ul},1)} \! \!= p_{k}  \widehat{\vect{h}}_{k}^{\Htran}\vect{D}_{k} \left( \sum\limits_{i=1,i\ne k}^{K} \!\!p_{i}  \vect{D}_{k}\widehat{\vect{h}}_{i} \widehat{\vect{h}}_{i}^{\Htran}\vect{D}_{k} + \vect{Z}_{k} \!\!\right)^{\!\dagger}  \!\! \vect{D}_{k}\widehat{\vect{h}}_{k}.\!\!
\end{align}
\end{corollary}
\begin{IEEEproof}
It follows from \cite[Lemma B.10]{massivemimobook} since \eqref{eq:uplink-instant-SINR} is a generalized Rayleigh quotient with respect to $\vect{v}_{k}$. To take into account that the matrix ${\bf{Z}}_k$ may not be strictly positive definite, a general form of the solution is used with $(\cdot)^{\dagger}$ denoting the matrix pseudoinverse \cite{Yu2007}.
\end{IEEEproof}

\begin{table*}[t]
            \caption{Number of complex multiplications for a generic UE $k$ of different combining schemes in each coherence block.}              \label{tab:cost_linear_processing} 
\centering
    \begin{tabular}{|c|c|c|c|} 
    \hline
     {Scheme} & {Channel estimation}&{Combining vector computation}   \\      \hline\hline 
    	 MMSE  &  $(N\tau_p + N^2)K |\mathcal{M}_k|$ &$\frac{\left(N|\mathcal{M}_k|\right)^2 + N|\mathcal{M}_k|}{2} K + \left(N|\mathcal{M}_k|\right)^2+\frac{\left(N|\mathcal{M}_k|\right)^3-N|\mathcal{M}_k|}{3} $\\   \hline
	     	 P-MMSE  &  $(N\tau_p + N^2) |\mathcal{P}_k| |\mathcal{M}_k|$  &$\frac{\left(N|\mathcal{M}_k|\right)^2 + N|\mathcal{M}_k|}{2} |\mathcal{P}_k| + \left(N|\mathcal{M}_k|\right)^2+\frac{\left(N|\mathcal{M}_k|\right)^3-N|\mathcal{M}_k|}{3} $\\   \hline
    	LP-MMSE   &   $(N\tau_p + N^2)\!\!\!\!\sum\limits_{l \in \mathcal{M}_k}\!\!|\mathcal{D}_l|\!\!$ &$\frac{N^2 + N}{2} \!\!\!\!\sum\limits_{l \in \mathcal{M}_k}\!\!|\mathcal{D}_l| + \big(\frac{N^3-N}{3} + N^2 \big)|\mathcal{M}_k| $\\ \hline
	MR 		& $(N\tau_p + N^2)|\mathcal{M}_k|$  &$-$\\   \hline
    \end{tabular}
\end{table*}

The SINR-maximizing combiner in \eqref{eq:MMSE-combining} minimizes the mean-squared error ${\rm{MSE}}_k = \mathbb{E} \{ | s_{k} - \widehat s_{k} |^2  \big| \{ \widehat{\vect{h}}_{i} \}  \}$, which is the conditional MSE between the data signal $s_k$ and the received signal $\widehat s_{k}$ in \eqref{eq:uplink-data-estimate} given the channel estimates \cite[Sec.~4.1]{massivemimobook}. This is why \eqref{eq:MMSE-combining} is called \emph{MMSE combining}. Notice that $ \vect{D}_{kl}\widehat{\vect{h}}_{il}=\vect{0}_N$ in \eqref{eq:MMSE-combining}  when $ \vect{D}_{kl}=\vect{0}_N$, which implies that we need to compute all the $K$ MMSE channel estimates $ \{\widehat{\vect{h}}_{il}: i =1,\ldots,K\}$ at any AP $l$ that is serving UE $k$ (i.e., APs with index $l\in \mathcal{M}_k$). The total number of complex multiplications required by MMSE combining is reported in Table~\ref{tab:cost_linear_processing} and unfortunately grows with $K$, thus making the complexity unscalable.

To solve this issue, we recall from \cite{Nayebi2016a} that the interference that affects UE $k$ is mainly generated by a small subset of the other UEs. Inspired by this, we propose that only the UEs that are served by partially the same APs as UE $k$ should be included in the expression in \eqref{eq:MMSE-combining}. These UEs have indices in the set
\begin{equation}
\mathcal{P}_k = \{ i: \vect{D}_{k} \vect{D}_{i} \neq \vect{0}_{LN} \}.
\end{equation}
Additional UEs can be included in $\mathcal{P}_k$ to deal with strongly interfering UEs that are only served by other APs, but such fine-tuning is outside the scope of this paper.

By utilizing $\mathcal{P}_k$, we propose an alternative \emph{partial MMSE (P-MMSE) combining} scheme:\footnote{Similar P-MMSE combining schemes have appeared in literature (e.g., \cite{Nayebi2016a}), but have not been designed for scalability.}
\begin{align} 
\vect{v}_{k}^{{\rm P-MMSE}} 
=    p_{k} \left( \sum\limits_{i \in \mathcal{P}_k} p_{i}  \vect{D}_{k} 
\widehat{\vect{h}}_{i} \widehat{\vect{h}}_{i}^{\Htran}
\vect{D}_{k} + {\bf{Z}}_k^\prime \right)^{\!\dagger} \!\!\vect{D}_{k}\widehat{\vect{h}}_{k} \label{eq:PMMSE-combining}
\end{align}
with 
\begin{align} 
{\bf{Z}}_k^\prime 
=  \vect{D}_{k}\left(\sum\limits_{i \in \mathcal{P}_k} p_{i}  
\vect{C}_{i}
+ \sigma_{\rm ul}^2  \vect{I}_{LN} \right)\vect{D}_{k}.
\end{align}
P-MMSE coincides with MMSE when UE $k$ is served by all APs, but is generally different. Note that $|\mathcal{P}_k|=\tau_p$ if all the APs that serve UE $k$ communicate with exactly the same set of UEs. Moreover, it holds that $|\mathcal{P}_k| \leq (\tau_p-1) |\mathcal{M}_k|+1$, where the upper bound is achieved in the unlikely case that all the APs in $\mathcal{M}_k$ serve UE $k$ but otherwise serve entirely different sets of UEs. Importantly, the upper bound is independent of $K$.
The total number of complex multiplications required by P-MMSE combining is reported in Table~\ref{tab:cost_linear_processing} and, as anticipated, this is a scalable scheme whose complexity does not grow with $K$.

\subsubsection{Distributed combining}
Instead of sending $\{ \vect{y}_{tl}^{\rm{pilot}}: t=1,\ldots,\tau_p\}$ and $\vect{y}_{l}^{\rm{ul}}$ to the CPU, AP $l$ can locally select the combiner $\vect{v}_{kl}$ on the basis of its local channel estimates $\{\widehat{\vect{h}}_{il}: i\in \mathcal D_{l}\}$, which are not more than $\tau_p$ vectors. The AP then computes its local estimate of $s_k$ as
\begin{align}
\widehat s_{kl} = \vect{v}_{kl}^{\Htran} \vect{D}_{kl} {\bf y}_{l}^{\rm{ul}}.
\end{align}
The local estimates of all APs that serve UE $k$ are then sent to a CPU where the final estimate of $s_k$ is obtained by taking the sum of the local estimates:
\begin{align}
\widehat s_{k} = \sum_{l=1}^{L} \widehat s_{kl}.
\end{align}
Since AP $l$ only needs to compute the local estimates for $|\mathcal{D}_l|$ UEs, $\tau_u|\mathcal{D}_l|$ complex scalars are sent to the CPU per coherence block. This number is upper bounded by $\tau_u \tau_p$, which does not grow with $K$ and therefore is scalable. The fronthaul signaling is summarized in Table~\ref{tab:signaling}.

Since the CPU does not have knowledge of channel estimates in the distributed case, an achievable UL SE cannot be computed as in Proposition \ref{theorem:uplink-capacity-general}. Instead, we utilize the so-called \emph{use-and-then-forget bound} that is widely used in Massive MIMO \cite[Th.~4.4]{massivemimobook}, and also in \cite{Nayebi2016a,Bashar2019a,Bjornson2019c} for Cell-Free Massive MIMO with $\vect{D}_i=\vect{I}_M \, \forall i$ and specific combining vectors.

\begin{proposition} \label{theorem:uplink-capacity-UatF}
An achievable UL SE for UE $k$ is 
\begin{align} \label{eq:uatf-ul} 
\mathacr{SE}_{k}^{({\rm ul},2)} = \frac{\tau_u}{\tau_c} \log_2  \left( 1 + \mathacr{SINR}_{k}^{({\rm ul},2)}  \right) 
\end{align}
where 
\begin{align} \notag
&\mathacr{SINR}_{k}^{({\rm ul},2)} = \\ &\frac{ p_{k} \left|  \mathbb{E} \left\{ \vect{v}_{k}^{\Htran} \vect{D}_{k} \vect{h}_{k} \right\} \right|^2  }{ 
\sum\limits_{i=1}^{K} p_{i} \mathbb{E}\!\Big\{ \! \Big|   \vect{v}_{k}^{\Htran} \vect{D}_{k}\vect{h}_{i} \Big|^2 \! \Big\} \!-\! p_{k} \Big|  \mathbb{E} \!\left\{ \vect{v}_{k}^{\Htran} \vect{D}_{k}\vect{h}_{k} \right\} \! \Big|^2 \!+\!\sigma_{\rm ul}^2 \mathbb{E}\{ \|  \vect{D}_{k}\vect{v}_{k}  \|^2\}}.\label{eq:uplink-instant-SINR-level2}
\end{align}
\end{proposition}
\begin{IEEEproof}
It follows the same approach as in \cite[Th.~4.4]{massivemimobook}, but for the received signal in \eqref{eq:uplink-data-estimate}.
\end{IEEEproof}

AP $l$ needs to select its combining vectors $ \vect{v}_{kl}$ for $k \in \mathcal{D}_l $ as a function of $\{\widehat{\vect{h}}_{il} : i \in \mathcal{D}_l \}$, without knowing the channel estimates available at other APs. Any combining vector $\vect{v}_{kl}$ that depends on the local channel estimates and statistics can thus be adopted in the above expression. The simplest solution is MR combining as in \cite{Ngo2017b,Nayebi2016a}, while it was shown in  \cite{Bjornson2019c} that L-MMSE combining provides better performance.
Unfortunately, L-MMSE is not a scalable scheme, but inspired by P-MMSE combining in \eqref{eq:PMMSE-combining}, we propose the local P-MMSE (LP-MMSE) given by
\begin{equation} \label{eq:MMSE-combining-single-AP-1}
\vect{v}_{kl}^{{\rm LP-MMSE}} =  p_{k}  \left( \sum\limits_{i \in \mathcal{D}_l} p_{i} \left( \widehat{\vect{h}}_{il} \widehat{\vect{h}}_{il}^{\Htran} + \vect{C}_{il} \right) + \sigma^2  \vect{I}_{N} \right)^{\!-1} \!\!   \widehat{\vect{h}}_{kl}.\!\!
\end{equation}
The key difference from L-MMSE is that \eqref{eq:MMSE-combining-single-AP-1} only includes the channel estimates and statistics of the UEs that AP $l$ is serving (i.e., those whose index $i \in \mathcal{D}_l$). The computational complexity of LP-MMSE is quantified in Table~\ref{tab:cost_linear_processing}. The fact that $|\mathcal{D}_l| \le \tau_p$ with $\tau_p$ being independent of $K$ makes LP-MMSE a scalable solution, as $K\to \infty$. 
Compared to centralized P-MMSE combining in \eqref{eq:PMMSE-combining}, LP-MMSE has much lower complexity per UE since it requires to compute the inverse of an $N\times N$, rather than $N|\mathcal{M}_k|\times N|\mathcal{M}_k|$, matrix.

The expectations in \eqref{eq:uplink-instant-SINR-level2} cannot be computed in closed form when using LP-MMSE, but can be easily computed using Monte Carlo simulations.
However, similar to \cite[Cor.~4.5]{massivemimobook}, we can obtain the following closed-form expression when using MR combining.

\begin{corollary} \label{cor:closed-form_ul}
If MR combining with $ \vect{v}_{kl} = \widehat{\vect{h}}_{kl}$ is used, then the expectations in \eqref{eq:uplink-instant-SINR-level2} become
\begin{equation} 
\mathbb{E} \left\{  \vect{v}_{k}^{\Htran} \vect{D}_{k} \vect{h}_{k} \right\}= p_k\tau_p{\sum_{l =1}^L \tr( \vect{D}_{kl} \vect{R}_{kl} \vect{\Psi}_{t_k l}^{-1} \vect{R}_{kl})}
\end{equation}
\begin{equation} 
 \mathbb{E} \left\{ \left\|  \vect{D}_{k} \vect{v}_{k} \right\|^2 
\right\}= {p_k \tau_p\sum_{l =1}^{L} \tr( \vect{D}_{kl} \vect{R}_{kl} \vect{\Psi}_{t_k l}^{-1} \vect{R}_{kl})}
\end{equation}
and
\begin{align}  \notag
&\mathbb{E} \big\{ \big| \vect{v}_{k}^{\Htran} \vect{D}_{k} \vect{h}_{i} \big|^2 \big\} = p_k\tau_p \sum_{l =1}^L \tr \left(  \vect{D}_{kl}\vect{R}_{il} \vect{R}_{kl} \vect{\Psi}_{t_k l}^{-1} \vect{R}_{kl}  \right) \\ &+ \begin{cases}
p_kp_i\tau_p^{2}  \left|\sum\limits_{l =1}^L\tr \left(  \vect{D}_{kl} \vect{R}_{il} \vect{\Psi}_{t_k l}^{-1} \vect{R}_{kl}  \right)\right|^2   &  \text{if }  t_i = t_k
\\
0 & \textrm{otherwise}
\end{cases}
\end{align}
where $t_i$ is the index of the pilot assigned to UE $i$.
\end{corollary}

\subsection{Downlink Data Transmission}
Next, we derive achievable DL SE expressions for the DCC signal model in \eqref{eq:DCC} and propose scalable precoding schemes. We use the \emph{hardening bound} that is widely used in the Massive MIMO literature to compute SEs \cite[Th.~4.6]{massivemimobook}, and also used in \cite{Ngo2017b,Nayebi2017a,Bjornson2019c} for Cell-Free Massive MIMO with $\vect{D}_i=\vect{I}_M \, \forall i$, for specific choices of precoding. Without loss of generality, we assume that
\begin{equation}
\vect{w}_{i} = \sqrt{ \rho_{i} }\bar{\vect{w}}_{i}  
\end{equation}
where $\bar{\vect{w}}_{i} \!\in\!\mathbb{C}^{N}$ determines the spatial directivity of the transmission and satisfies ${\mathbb{E} \{ \| \bar{\vect{w}}_{i} \|^2 \}} = 1$ such that $\rho_{i}\!\geq\!0$ is the total transmit power allocated to UE $i$. 

\begin{proposition}\label{theorem:downlink-capacity-hardening}
An achievable DL SE for UE $k$ is given by
\begin{equation} \label{eq:downlink-rate-expression-general}
\mathacr{SE}_{k}^{({\rm dl})} = \frac{\tau_d}{\tau_c} \log_2  \left( 1 + \mathacr{SINR}_{k}^{({\rm dl})}  \right) 
\end{equation}
where 
\begin{align}
\mathacr{SINR}_{k}^{({\rm dl})} &\!=\! \frac{ \rho_k\big|   \mathbb{E} \left\{  \vect{h}_{k}^{\Htran} \vect{D}_{k} \bar{\vect{w}}_{k} \right\} \big|^2 }{ \sum\limits_{i=1}^{K} \rho_i\mathbb{E} \big\{ \big|  \vect{h}_{k}^{\Htran} \vect{D}_{i} \bar{\vect{w}}_{i} \big|^2 \big\} -  \rho_k\big|  \mathbb{E} \left\{  \vect{h}_{k}^{\Htran} \vect{D}_{k} \bar{\vect{w}}_{k} \right\} \big|^2 + \sigma_{\rm dl}^2 } \label{eq:downlink-SINR-level1}
\end{align}
and the expectations are with respect to the channel realizations.
\end{proposition}
\begin{IEEEproof}
It follows the same approach as in \cite[Th.~4.6]{massivemimobook}, but for the signal model in \eqref{eq:DCC}.
\end{IEEEproof}
The DL SE of UE $k$ depends on the normalized precoding vectors of all UEs (i.e., $\{\bar{\vect{w}}_i : i=1,\ldots,K \}$) in contrast to the UL SEs in Propositions \ref{theorem:uplink-capacity-general} and \ref{theorem:uplink-capacity-UatF} that only depend on the UE's own combining vector ${\bf v}_k$. Hence, while receive combining can be optimized on a per-UE basis, the precoding vectors should ideally be optimized jointly for all UEs, which is not scalable. To obtain a good heuristic solution, we utilize the following UL-DL duality result.

\begin{proposition} \label{prop:duality}
Let $\{{\bf v}_i: i=1,\ldots,K \}$ and $\{p_i: i=1,\ldots,K \}$  denote the set of combining vectors and transmit powers used in the UL. If the normalized precoding vectors are selected as 
\begin{equation}\label{eq:precoding-vectors-duality-proposition}
\bar{\vect{w}}_{i}  = \frac{{\bf v}_{i}}{\sqrt{\mathbb{E} \{ {\vect{v}}_{i}^{\Htran}\vect{D}_{i} {\vect{v}}_{i}\}}} 
\end{equation}
then there exists a DL power control policy $\{\rho_i:\forall i\}$
with $\sum_{i=1}^{K} \rho_i/\sigma^{2}_{\rm dl}=\sum_{i=1}^{K}p_i/\sigma^{2}_{\rm ul}$ for which 
\begin{align}
\mathacr{SINR}_{k}^{({\rm dl})} = \mathacr{SINR}_{k}^{({\rm ul},2)}\quad \forall k.
\end{align}
\end{proposition}

\begin{IEEEproof}
This is proved by following the same approach as in \cite[Th.~4.8]{massivemimobook}, but for the signal model in \eqref{eq:DCC}. Details are given in the Appendix for completeness. Notice that the total transmit power in DL is the same as in UL, but is allocated differently over the UEs. The exact expression for the power control coefficients is given in the Appendix.
\end{IEEEproof}
This theorem shows that the SINRs that are achieved in the UL are also achievable in the DL, by selecting the power control coefficients $\{\rho_i:\forall i\}$ and normalized precoding vectors $\{\bar{\vect{w}}_i :\forall i\}$ properly. Consequently, we have that an achievable DL SE for UE $k$ is $\mathacr{SE}_{k}^{({\rm dl})} = \frac{\tau_d}{\tau_c} \log_2  ( 1 + \mathacr{SINR}_{k}^{({\rm ul},2)}  ).$
Despite the UL-DL duality, the SE usually differs in UL and DL since the required power control policy has unscalable complexity and the power constraints in UL and DL also prevent it from being fully used.
However, Proposition~\ref{prop:duality} motivates to select the DL precoders in cell-free networks based on the UL combiners as in \eqref{eq:precoding-vectors-duality-proposition}.\footnote{This is consistent with cellular networks, where it is a common practice to let the DL precoders be equal to the UL combiners, except for a scaling factor \cite{Boche2002a,Viswanath2003a,Bjornson2016a}.}
This must be done on the basis of the available channel estimates and, importantly, if the UL selection is scalable this property carries over to the DL. As in the UL, we consider two levels of cooperation among APs for precoding design. We stress that since
\begin{align}
\begin{cases}
\vect{D}_{il} \vect{w}_{il}=\vect{w}_{il} & i \in \mathcal{D}_l \\
\vect{D}_{il} \vect{w}_{il} = \vect{0}_N & i \not \in \mathcal{D}_l
\end{cases}
\end{align}
only the precoding vectors $ \vect{w}_{il}$ for $i \in \mathcal{D}_l $ need to be selected for AP $l$. At both levels, we assume that the APs delegate the task of DL data encoding to a nearby CPU.

\begin{remark}
The hardening bound of Proposition \ref{theorem:downlink-capacity-hardening} has been widely used since the early articles on Massive MIMO \cite{Marzetta2010a,Hoydis2013,ngo2013energy} and is a reasonable choice for channels that exhibit \emph{channel hardening} \cite[Sec. 2.5]{massivemimobook}. However, the bound can be rather loose in Cell-Free Massive MIMO when the number of antennas $N$ at the APs is relatively small \cite{Chen2018b}, but it depends on the choice of combining scheme. We will return to this potential issue in Section~\ref{sec:numerical}.
\end{remark}

\subsubsection{Centralized precoding}
At the first level, the CPU uses the UL channel estimates to compute the normalized precoding vectors $\{\bar{\vect{w}}_{il}\}$ by exploiting channel reciprocity. Motivated by the UL-DL duality, we select the DL precoding vectors according to \eqref{eq:precoding-vectors-duality-proposition}. By choosing $\vect{v}_i$ according to one of the UL combining schemes described earlier in Section~\ref{centralized_combining_UL}, the corresponding precoding scheme is obtained; that is, $\vect{v}_i = \vect{v}_i^{\textrm{P-MMSE}}$ yields P-MMSE precoding, and so forth. Once the precoding vectors are computed, they are used by the CPU to form the DL signal of any given AP $l$ 
\begin{equation}
\vect{x}_{l}^{\rm{dl}} = \sum_{i=1}^{K} \sqrt{ \rho_{i} } \vect{D}_{il}  \bar{\vect{w}}_{il} \varsigma_i
\end{equation}
which is sent to the AP via the fronthaul link for transmission.

The signaling required can be quantified as follows. In each coherence block, AP $l$ needs to send $\tau_pN$ complex scalars to the CPU representing the pilot signals and to receive $\tau_dN$ complex scalars from the CPU for the DL signals. These values are summarized in Table~\ref{tab:signaling}. 

\subsubsection{Distributed precoding} Instead of sending $\{ \vect{y}_{tl}^{\rm{pilot}}: t=1,\ldots,\tau_p\}$ to the CPU, AP $l$ can locally select the precoding vector $\bar{\vect{w}}_{il}$ on the basis of its local channel estimates $\{\widehat{\vect{h}}_{il}: i\in \mathcal D_{l}\}$ to achieve a scalable implementation \cite{Bjornson2010c}. In this case, only the DL data signals $\{\varsigma_i : i \in \mathcal D_{l}\}$ are sent from the CPU to AP $l$ in each coherence block. This means that a total of $\tau_d|\mathcal{D}_l|$ complex scalars exchanged per coherence block, as summarized in Table~\ref{tab:signaling}. 

Two possible precoding schemes that satisfy the scalability requirement are the classical MR and the new LP-MMSE given in \eqref{eq:MMSE-combining-single-AP-1}. 
Note that MR is also known as conjugate beamforming and is the standard scheme in the Cell-Free Massive MIMO literature, while this is the first time that LP-MMSE
precoding is considered.\footnote{A variation on LP-MMSE precoding, known as signal-to-leakage-and-noise ratio (SLNR) precoding, was considered in the conference version \cite{BjornsonPIMRC2019}. It is another heuristic way to select precoding vectors in the absence of a UL-DL duality result. We refer to \cite{BjornsonPIMRC2019} for details.} The benefit of this scheme over MR is two-fold: 1) it suppresses interference spatially if $N>1$ since the vector maximizes the ratio between desired signal power and interference caused to the other UEs served by the same AP; and 2) it reduces the variations in the effective  gain $\vect{h}_{kl}^{\Htran}\vect{D}_{il}\vect{w}_{il}$ of desired and interfering channels for any $N$.

\begin{figure}[t!]
	\centering 
	\begin{overpic}[width=\columnwidth,tics=10]{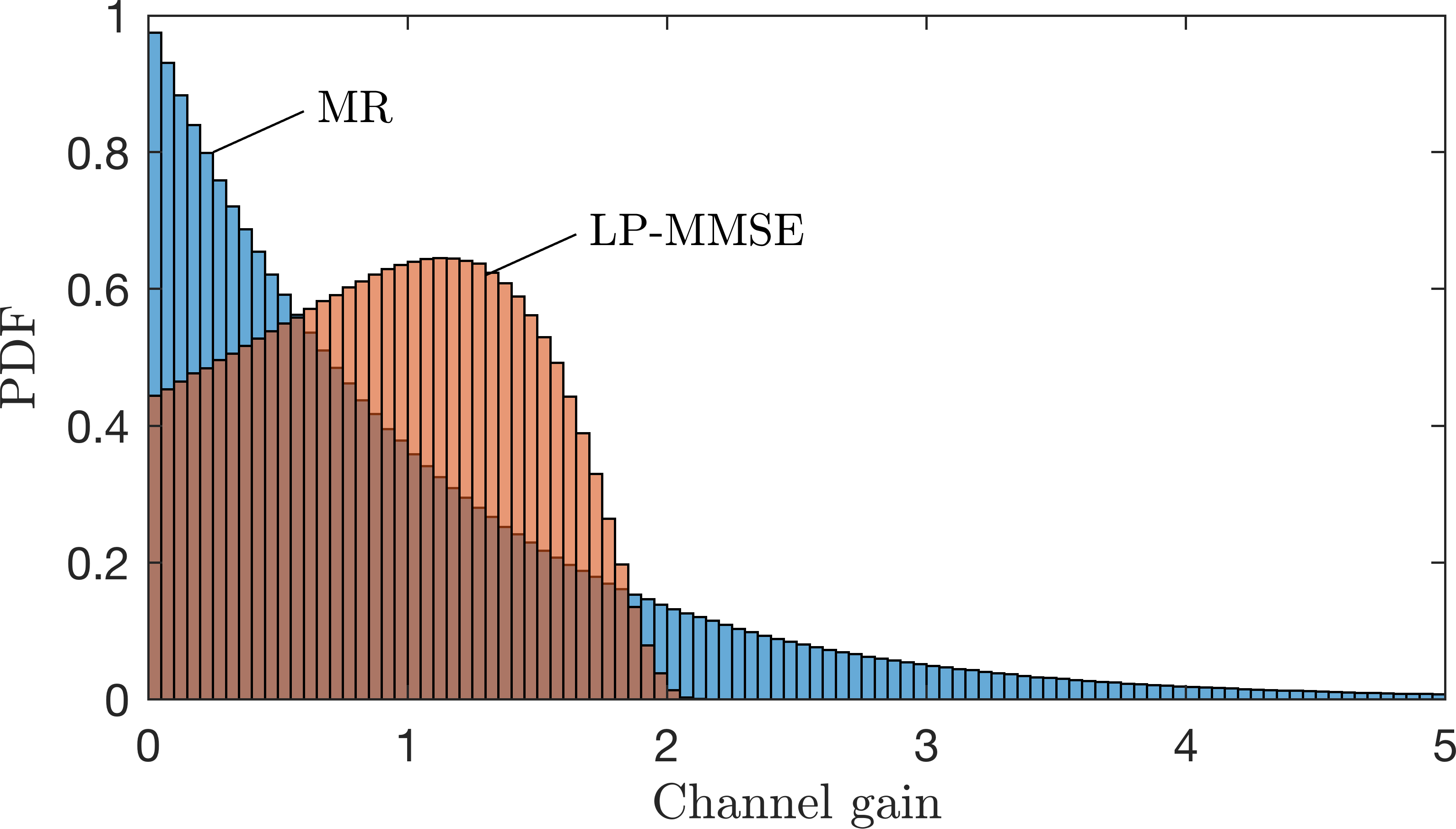}
\end{overpic} 
	\caption{For $N=L=K=\rho=\sigma_{\rm dl}^2=1$ and perfect CSI, the channel gain is $|h|^2$ with MR and $\frac{|h|^2}{|h|^2+1} / \sqrt{\mathbb{E} \{ \frac{|h|^2}{(|h|^2+1)^2}  \}}$ with LP-MMSE, where $x \sim \CN(0,1)$. Their PDFs are widely different, particularly since only LP-MMSE has bounded support while MR has not.}
	\label{fig:simulationGain}  
\end{figure}

The latter is a non-trivial phenomenon that appears even with $N=L=K=1$ and when perfect CSI is available. Fig.~\ref{fig:simulationGain} shows the probability density function (PDF) of the channel gains $h^{\Htran} w$ for $h \sim \CN(0,1)$ and $\rho=\sigma_{\rm dl}^2 =1$. We consider MR with $w = h$ and LP-MMSE with $w= \frac{h}{|h|^2+1} / \sqrt{ \mathbb{E} \{ \frac{|h|^2}{(|h|^2+1)^2}  \}}$.
The channel gains have roughly the same mean values $\mathbb{E} \{  h^{\Htran} w \}$, but MR gives an exponential distribution with an infinite tail while LP-MMSE has a small and compact support. This behavior will lead to higher SE when using LP-MMSE 
under inter-user interference; see Section~\ref{subsec:numerical-downlink} for details.

As in the UL, the SE can be computed in closed form with MR precoding, following the same approach as in \cite[Cor.~4.7]{massivemimobook}.

\begin{corollary} \label{cor:closed-form}
With MR precoding, the expectations in \eqref{eq:downlink-SINR-level1} become
\begin{equation} 
\mathbb{E} \{ \vect{h}_k^{\Htran} \vect{D}_k \vect{w}_k \} = \sum_{l=1}^L \sqrt{\rho_{il} p_k \tau_p \tr( \vect{D}_{kl} \vect{R}_{kl} \vect{\Psi}_{t_k l}^{-1} \vect{R}_{kl})}
\end{equation}
\begin{align} \notag
&\mathbb{E} \{ |\vect{h}_k^{\Htran} \vect{D}_i \vect{w}_i|^2 \}  = \sum_{l=1}^{L} \rho_{il} \frac{  \tr \left(  \vect{D}_{il} \vect{R}_{il} \vect{\Psi}_{t_i l}^{-1} \vect{R}_{il} \vect{D}_{il} \vect{R}_{kl}  \right) }{ \tr( \vect{R}_{il} \vect{\Psi}_{t_i l}^{-1} \vect{R}_{il})} \\ &+ \begin{cases}
\left|  \sum\limits_{l=1}^{L} \sqrt{ \rho_{il} p_k \tau_p} \frac{  \tr \left(  \vect{D}_{il} \vect{R}_{il} \vect{\Psi}_{t_i l}^{-1} \vect{R}_{kl}  \right) }{ \sqrt{ \tr( \vect{R}_{il} \vect{\Psi}_{t_i l}^{-1} \vect{R}_{il}) } } \right|^2 &  \text{if }  t_i = t_k
\\
0 & \textrm{otherwise}
\end{cases}
\end{align} 
where $t_i$ is the index of the pilot assigned to UE $i$.
\end{corollary}

\subsection{Uplink and Downlink Power Allocation}
\label{subsec:power-allocation}

The UL transmit powers $\{ p_k :  k=1,\ldots,K \}$ in \eqref{eq:uplink-rate-expression-general} and \eqref{eq:uatf-ul} need to be selected. The network-wide power optimization methods in \cite{Ngo2017b,Nayebi2016a,Bashar2019a} are not scalable, thus a heuristic solution is needed. We assume that each UE has a maximum UL power of $P$. 
Since heuristic solutions must be fine-tuned based on extensive measurements, which is outside the scope of this paper, we consider a good baseline scheme. Each UE transmits at full power, which was shown in  \cite{Bjornson2019c} to provide good SE for both strong and weak UEs:
\begin{equation}
p_i = P\quad \quad \text{for}\quad i = 1,\ldots, K.
\end{equation}

Similarly, each AP has a maximum transmit power, denoted by $\rho$, and needs to select how to allocate it between the UEs it is serving.
Network-wide optimization algorithms, as developed in \cite{Bjornson2013d,Ngo2017b,Nayebi2017a}, are not scalable as $K \to \infty$ since the number of optimization variables grows with $K$.\footnote{It is possible to implement network-wide optimization problems in an iterative semi-distributed way, for example, using dual decomposition theory \cite[Sec.~4.3]{Bjornson2013d}. However, these approaches converge slowly, require even more optimization variables, and require a lot of backhaul signaling. Hence, this approach is neither practical nor scalable.} Since each AP is (at least partially) unaware of the power allocation decisions made at other APs, only heuristic solutions are scalable. There are plenty of such schemes in the literature; some examples
 are found in \cite{Bjornson2010c,Bjornson2011a,Nayebi2017a,Interdonato2019a}, \cite[Sec.~3.4.4]{Bjornson2013d}. Evaluation and comparison of the existing heuristic schemes require extensive simulations, which is outside the scope of this paper. Hence, we have selected schemes that are known to work fairly well.
 
 For the centralized precoding, we consider simple network-wide equal power allocation with $\rho_i = \frac{\rho}{\tau_p}$. The unit-norm precoding vector $\bar{\vect{w}}_{i}$ determines how this power is distributed between the APs, and all APs are guaranteed to satisfy their power constraint since they serve at most $\tau_p$ UEs with a per-UE power that is at most $\frac{\rho}{\tau_p}$.

For the distributed precoding, we  adopt the power allocation algorithm from \cite{Interdonato2019a}:
 \begin{equation} \label{eq:equal_power_DL}
\rho_{kl} = \begin{cases} \rho \frac{\sqrt{\beta_{kl}}}{\sum_{i \in \mathcal{D}_l} \sqrt{\beta_{il}}} & \textrm{if } k \in \mathcal{D}_l\\
0 &  \textrm{otherwise}
\end{cases}
\end{equation}
where $\beta_{kl} = \tr(\vect{R}_{kl})/N$. Note that both schemes have the feature that each AP allocates more power to UEs with strong channels than to UEs with weak channels, while guaranteeing non-zero powers to all the served UEs.
Since each UE is served by at least one AP (i.e., its Master AP), it will be assigned non-zero transmit power when using \eqref{eq:equal_power_DL} and, thus, get a non-zero SE.

\subsection{Network Topology} \label{subsec:network_topology}

The proposed algorithms have been described as if the network has a star topology, where the APs are connected to a single CPU. However, they are mostly transparent to the actual network topology since only neighboring APs cooperate, which means that many other implementations are possible. Importantly, the CPU should not be viewed as a physical unit, but rather as a set of centralized processing tasks that must be carried out somewhere in the network.

One option is to have local processors at each AP, as in classic cellular networks, and backhaul connections to the core network. There is no physical CPU in this case, but its tasks are divided between the APs \cite{Bjornson2013d}; for example, the Master AP of UE $k$ can be the one taking care of the CPU tasks that are related to this UE (e.g., encoding of DL data and decoding of UL data). When information is sent to/from the CPU, it is actually sent to the AP that is responsible for carrying out the related CPU tasks, which might be a different AP over time.

Another option is to divide the APs into disjunct sets and connect each one via fronthaul to an edge-cloud processor \cite{Perlman2015a,Burr2018a,Interdonato2019a} for centralized processing, as illustrated in Fig.~\ref{fig:cell-free}. The CPU tasks are distributed between different physical edge-cloud processors in this case. 

There are many other cloud-RAN solutions that can be used to distribute the computations over the network; see \cite{6882182,Peng2016a} for some examples.

\section{Numerical Analysis}
\label{sec:numerical}

Numerical results are used in this section to demonstrate that the proposed way to make Cell-Free Massive MIMO scalable leads to a negligible loss in performance. We consider a simulation scenario where $M$ APs and $K=100$ UEs are independently and uniformly distributed in a $2 \times 2$\,km square. Two different setups are considered: $i$) $L=400$ APs with $N=1$ antenna; $ii$) $L=100$ APs with $N=4$ antennas. By using the wrap-around technique, we approximate an infinitely large network with 100 antennas/km$^2$ and 25 UEs/km$^2$. 

The UEs connect to the network as described in Section~\ref{eq:pilot-assignment}, starting with $\tau_p$ UEs that have different pilots and then letting the UEs connect one after the other. We use the same propagation model as in \cite[Sec.~4.1.3]{massivemimobook} with spatially correlated fading. The only difference is that the APs are deployed 10\,m above the UEs, which creates a natural minimum distance. We assume $\tau_c = 200$, $\tau_p=10$, $p_k=100$\,mW, $\rho=1$\,W, and 20\,MHz bandwidth. We use $\tau_u = 190$ and $\tau_d = 190$ when evaluating UL and DL, respectively.

\subsection{Uplink}

\begin{figure}
        \centering 
        \begin{subfigure}[b]{\columnwidth} \centering 
                \includegraphics[width=\textwidth, trim=13 0 25 0, clip]{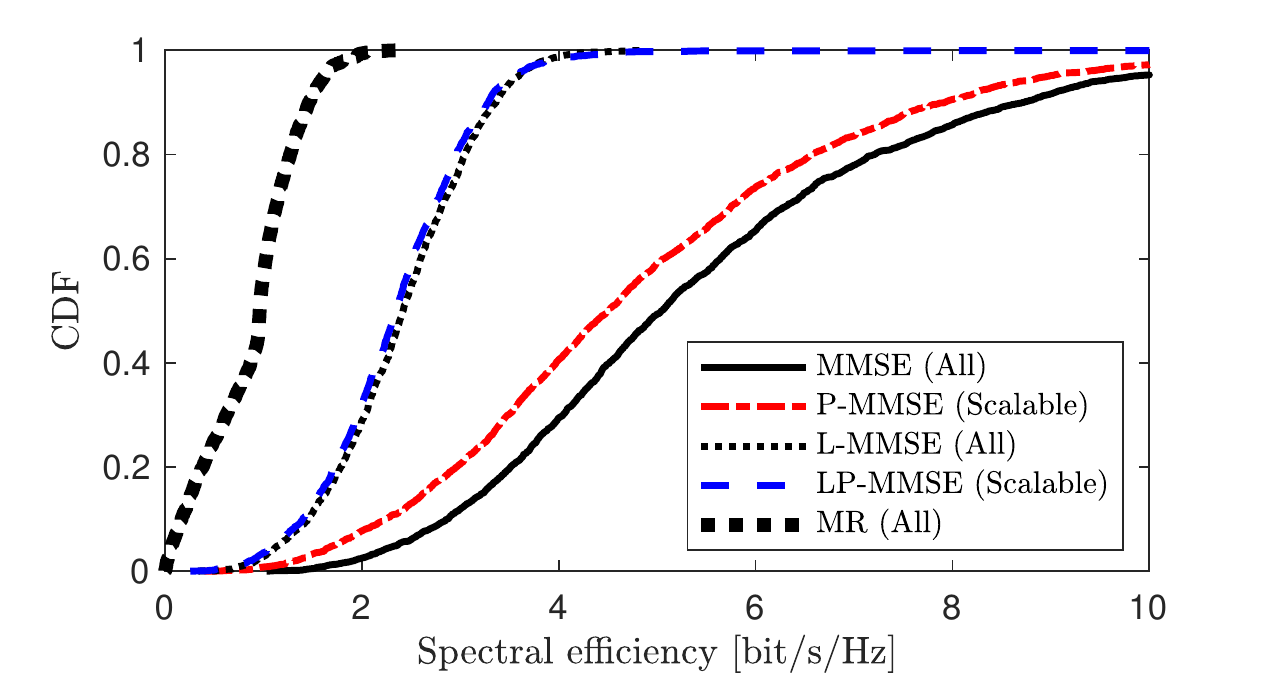}  \vspace{-2mm}
                \caption{$L=400$ APs with $N=1$ antenna.} 
                \label{fig:simulationSE_N1_uplink}
        \end{subfigure} 
        \begin{subfigure}[b]{\columnwidth} \centering
                \includegraphics[width=\textwidth, trim=13 0 25 0, clip]{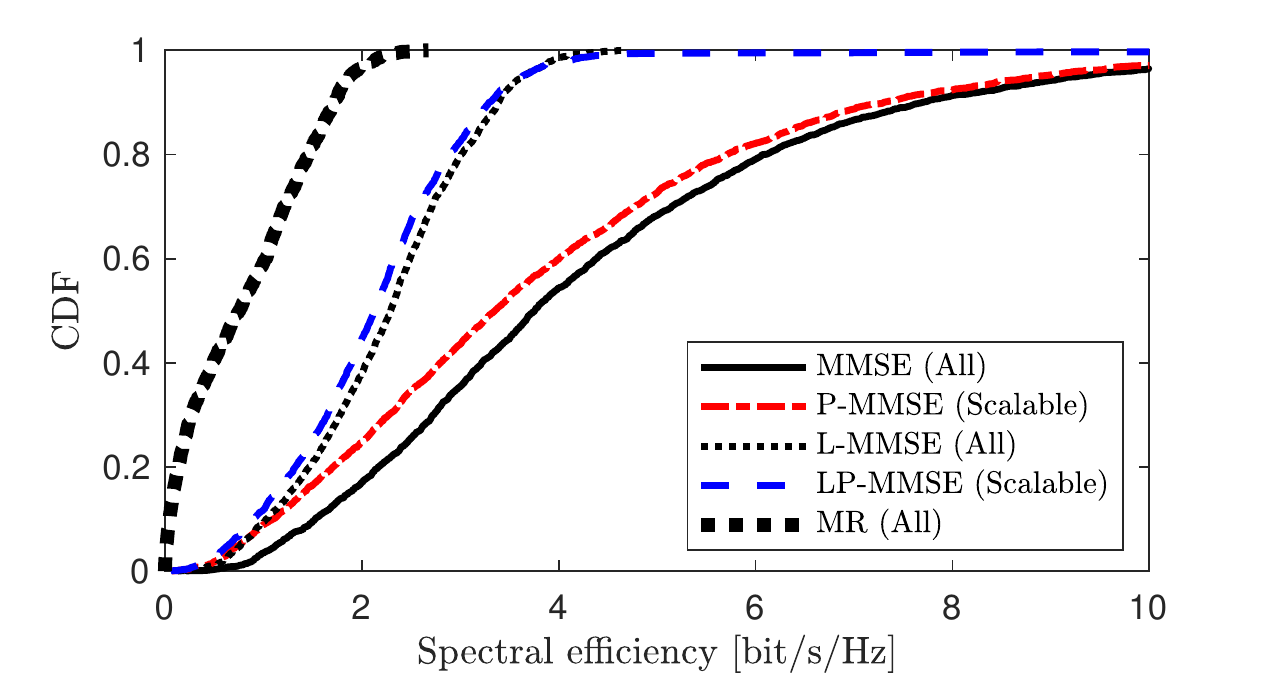} \vspace{-2mm}
                \caption{$L=100$ APs with $N=4$ antennas.}  
                \label{fig:simulationSE_N4_uplink} 
        \end{subfigure}  
\vspace{-2mm}
        \caption{UL SE per UE with different scalable and non-scalable ``(All)'' combining schemes.} 
        \label{fig:simulationSE_uplink} 
\end{figure}

Fig.~\ref{fig:simulationSE_uplink} shows the cumulative distribution function (CDF) of the UL SE per UE. We compare the proposed scalable centralized P-MMSE and distributed LP-MMSE schemes with three benchmarks where all APs serve all UEs: optimal centralized MMSE combining in \eqref{eq:MMSE-combining} from \cite{Nayebi2016a,Bjornson2019c}, distributed L-MMSE combining from \cite{Bjornson2019c}, and conventional MR from the original paper \cite{Ngo2017b} on the topic. These three are marked with ``(All)'' and we stress that none of these benchmarks is scalable, according to Definition \ref{def:scalable}.
The first observation is that the proposed distributed LP-MMSE performs very well; the average SE is $2.7\times$ higher than with MR and the performance loss compared to L-MMSE is negligible. When it comes to the two centralized schemes, the proposed scalable P-MMSE achieves 89\% of the average SE with optimal MMSE combining. The performance loss comes from two factors: limiting the number of APs that serves each UE and using the proposed scalable, but suboptimal, algorithm for cluster formation. Since the loss is small, the price to pay for scalability is also small and the algorithm performs well. The intuition is that the few closest APs receive the vast majority of the total received power for a given UE, and these are the APs that our clustering algorithm selects to serve that UE. It is thus sufficient to suppress interference between the UEs that these APs are co-serving.

When comparing the two setups ($L=400,N=1$ and $L=100,N=4$), we notice that it is preferable to have many single-antenna APs. The UEs with the lowest SEs benefit the most from having many APs, while the most fortunate UEs achieve roughly the same SE when having fewer multi-antenna APs, thanks to the more capable local interference mitigation at the APs.

\begin{figure} 
        \centering 
        \begin{subfigure}[b]{\columnwidth} \centering 
                \includegraphics[width=\textwidth, trim=13 0 25 0, clip]{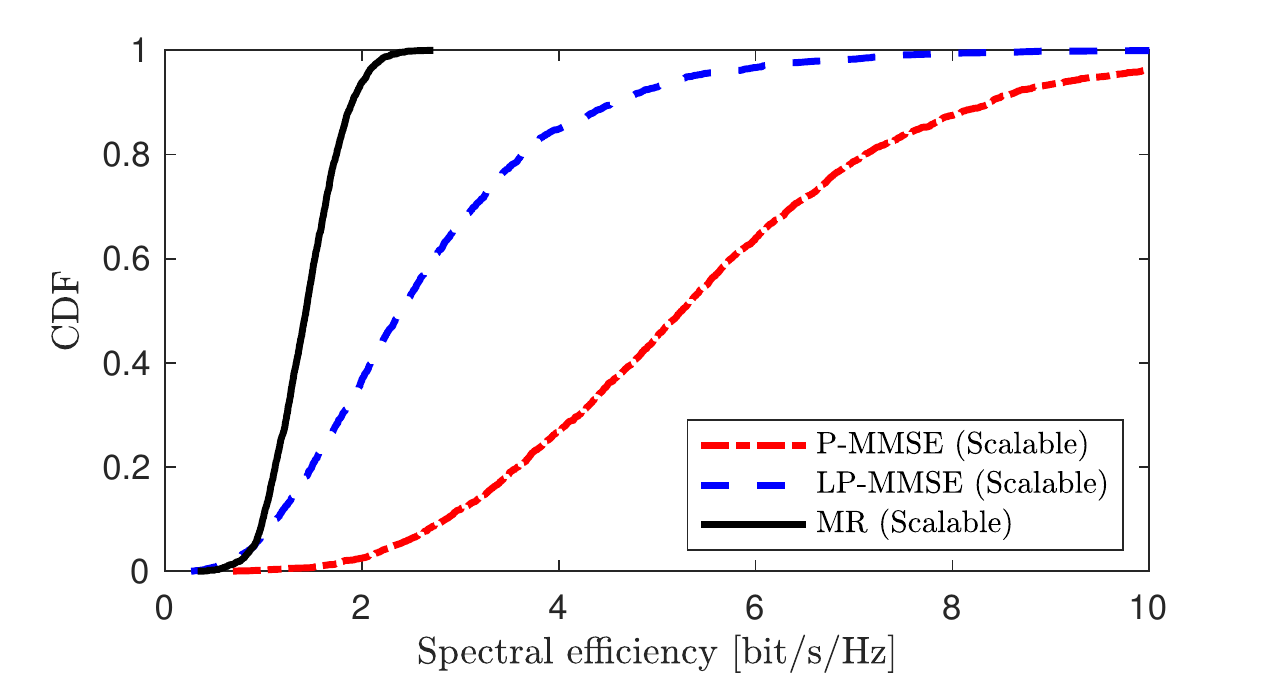}  \vspace{-2mm}
                \caption{$L=400$ APs with $N=1$ antenna.} 
                \label{fig:simulationSE_N1}
        \end{subfigure} 
        \begin{subfigure}[b]{\columnwidth} \centering
                \includegraphics[width=\textwidth, trim=13 0 25 0, clip]{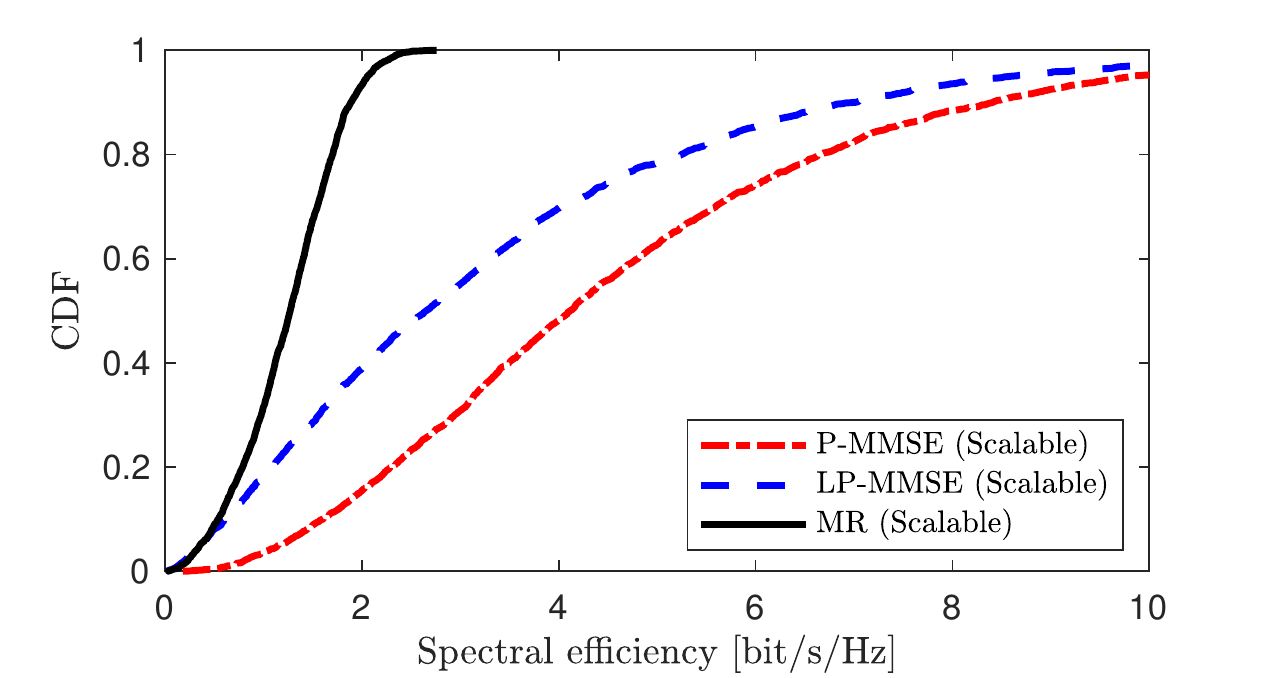} \vspace{-2mm}
                \caption{$L=100$ APs with $N=4$ antennas.}  
                \label{fig:simulationSE_N4} 
        \end{subfigure} \vspace{-2mm}
        \caption{DL SE per UE with different scalable precoding schemes.} 
        \label{fig:simulationSE_downlink} 
\end{figure}

\subsection{Downlink}
\label{subsec:numerical-downlink}

Since we have established that the proposed scalable Cell-Free Massive MIMO provides very competitive performance, the DL simulations will focus on two other aspects: interference-limited operation and tightness of the capacity lower bounds.

Fig.~\ref{fig:simulationSE_downlink} shows the CDF of the DL SE per UE for the two setups ($L=400,N=1$ and $L=100,N=4$). We compare three scalable precoding solutions: centralized P-MMSE, distributed LP-MMSE, and distributed MR (i.e., a combination of conventional MR with the proposed scalable clustering). The power allocation is selected as described in Section~\ref{subsec:power-allocation}, where the centralized scheme limits its power to guarantee that all APs satisfy their power constraints. As a consequence, the distributed precoding schemes use 40 times more transmit power, but anyway give lower SEs in Fig.~\ref{fig:simulationSE_downlink}. The reason is that the cell-free system is interference-limited, so the spatial interference mitigation enabled by centralized precoding is very beneficial.
The performance gap between P-MMSE and LP-MMSE is large in Fig.~\ref{fig:simulationSE_downlink}(a), but shrinks in Fig.~\ref{fig:simulationSE_downlink}(b) where each AP has $N=4$ antennas and thus can spatially suppress its interference. The different power allocation is also a reason for the performance gap.

Among the distributed schemes, LP-MMSE outperforms MR for 95\% of the UEs and gives comparable SE for the 5\% most unfortunate UEs.
A partial reason for this result is the different channel gain distributions that were illustrated in Fig.~\ref{fig:simulationGain}, where we recall that MR has much larger variations and therefore poorer tightness of the capacity bound in Proposition~\ref{theorem:downlink-capacity-hardening}. To verify this, we computed the SE in a genie-aided case where the UEs have perfect CSI. In the setup with $N=1$, the average SE with LP-MMSE is 90\% of the genie-aided case, while MR only achieves 60\%. The latter leads to limited channel hardening, as previously noted in \cite{Chen2018b}, and calls for alternative methods for DL channel estimation. However, a more convenient solution is to use LP-MMSE precoding, for which Proposition~\ref{theorem:downlink-capacity-hardening} is a tight capacity bound. Finally, we notice that P-MMSE achieved 98\% of the genie-aided case.

\section{Conclusions}\label{sec:conclusions}

This paper developed a new framework for scalable Cell-Free Massive MIMO systems, where the complexity and signaling at each AP is finite even when the number of UEs goes to infinity. To achieve this, we exploited concepts from the DCC framework previously used in the Network MIMO literature. We developed new scalable algorithms for initial access, pilot assignment, cooperation cluster formation, and both centralized and distributed signal processing for receive combining and transmit precoding. We demonstrated that MR (conjugate beamforming) is outperformed by the proposed distributed LP-MMSE combining/precoding, which in turn is outperformed by the centralized P-MMSE combining/precoding. Importantly, for a given power allocation policy, the scalability can be achieved with a negligible performance loss. The reason is that each UE receives most of its power from a small subset of the APs, due to the large pathloss variations, and this subset is identified by the proposed algorithms.
When the proposed precoding is used, the downlink capacity bounds are tight in cell-free networks, thus the tightness issue observed in \cite{Chen2018b} is mainly a problem for MR precoding.

While the proposed methods are nearly optimal, one aspect was not considered in detail: power allocation for centralized and distributed operation. Although there are plenty of scalable, heuristic algorithms, it is unknown how well they perform compared to centralized optimization.

\section*{Appendix}
Let $\gamma_{k} = \mathacr{SINR}_{k}^{({\rm ul},2)}$ denote the value of the effective SINR in \eqref{eq:uplink-instant-SINR-level2} for the UL powers $\{p_i: \forall i\}$ and combiners $\{{\bf v}_{i}: \forall i\}$. We want to show that $\gamma_{k} = \mathacr{SINR}_{k}^{({\rm dl})}$ is achievable in the DL when \eqref{eq:precoding-vectors-duality-proposition} is satisfied for all $i$. Plugging \eqref{eq:precoding-vectors-duality-proposition} into \eqref{eq:downlink-SINR-level1} yields the following SINR constraints:
\begin{align} 
\gamma_k  =
&\frac{{\rho_{k}}\bigg|    \mathbb{E} \left\{  \frac{\vect{h}_{k}^{\Htran} \vect{D}_{k} {\bf v}_{k}}{\sqrt{\mathbb{E} \{ \| {\vect{v}}_{k} \|^2 \}}}  \right\} \bigg|^2 }{ \sum\limits_{i=1}^{K} {\rho_{i}}\mathbb{E} \bigg\{ \bigg|   \frac{\vect{h}_{k}^{\Htran} \vect{D}_{i} {\bf v}_{i}}{\sqrt{\mathbb{E} \{ \| {\vect{v}}_{i} \|^2 \}}}   \bigg|^2 \bigg\} - {\rho_{k}} \bigg|   \mathbb{E} \left\{  \frac{\vect{h}_{k}^{\Htran} \vect{D}_{k} {\bf v}_{k}}{\sqrt{\mathbb{E} \{ \| {\vect{v}}_{k} \|^2 \}}}  \right\} \bigg|^2 + \sigma_{\rm dl}^2 }.\label{eq:downlink-SINR-level2-AppendixA-1}
\end{align}
We call $\boldsymbol{\Gamma} \in \mathbb{C}^{K \times K}$ a diagonal matrix with the $k$th diagonal element being
\begin{align}
[\boldsymbol{\Gamma}]_{kk} =\frac{1}{\gamma_k}\bigg|    \mathbb{E} \left\{ \vect{h}_{k}^{\Htran} \vect{D}_{k} \frac{{\bf v}_{k}}{\sqrt{\mathbb{E} \{ {\vect{v}}_{k}^{\Htran}\vect{D}_{k} {\vect{v}}_{k} \}}}  \right\} \bigg|^2
\end{align}
and let $\boldsymbol{\Sigma} \in \mathbb{C}^{K \times K}$ be the matrix whose $(i,k)$th element is
\begin{align}
[\boldsymbol{\Sigma}]_{ki} = \mathbb{E} \bigg\{ \bigg|  \vect{h}_{k}^{\Htran} \vect{D}_{i} \frac{{\bf v}_{i}}{\sqrt{\mathbb{E} \{ {\vect{v}}_{i}^{\Htran}\vect{D}_{i} {\vect{v}}_{i} \}}}   \bigg|^2 \bigg\} - \begin{cases}
0 & i \ne k \\
 \gamma_k[\boldsymbol{\Gamma}]_{kk}& i=k.
\end{cases}
\end{align}
Therefore, we may rewrite \eqref{eq:downlink-SINR-level2-AppendixA-1} as
\begin{align}
[\boldsymbol{\Gamma}]^{-1}_{kk} = {\rho_k}/({ \sum\limits_{i=1}^{K}\rho_i [\boldsymbol{\Sigma}]_{ki} + \sigma_{\rm dl}^2 })
\end{align}
from which we obtain $\sigma_{\rm dl}^2 = \rho_k [\boldsymbol{\Gamma}]_{kk} - \sum\nolimits_{i=1}^{K}\rho_i [\boldsymbol{\Sigma}]_{ki}$.
The $K$ constraints can be written in matrix form as $
{\bf 1}_K\sigma_{\rm dl}^2 = \left(\boldsymbol{\Gamma} - \boldsymbol{\Sigma}\right)\boldsymbol{\rho}$ 
with $\boldsymbol{\rho} =[\rho_1,\ldots,\rho_K]^T$ being the downlink transmit power vector. The SINR constraints are thus satisfied if 
\begin{align}\label{eq:optimal-power-appendix}
\boldsymbol{\rho} = \left(\boldsymbol{\Gamma} - \boldsymbol{\Sigma}\right)^{-1}{\bf 1}_K\sigma_{\rm dl}^2.
\end{align}
This is a feasible power if $\boldsymbol{\Gamma} - \boldsymbol{\Sigma}$ is invertible, which always holds when ${\bf p}$ is feasible. To show this, we notice that the $K$ UL SINR conditions can be expressed in a similar form where $\boldsymbol{\Sigma}$ is replaced by $\boldsymbol{\Sigma}^{\Ttran}$
such that ${\bf p} = \left(\boldsymbol{\Gamma} - \boldsymbol{\Sigma}^{\Ttran}\right)^{-1}{\bf 1}_K\sigma_{\rm ul}^2$.
Since the eigenvalues of $\boldsymbol{\Gamma} - \boldsymbol{\Sigma}$ and $\boldsymbol{\Gamma} - \boldsymbol{\Sigma}^{\Ttran}$ are the same and the UL SINR conditions are satisfied by assumption, we can always select the DL powers according to \eqref{eq:optimal-power-appendix}. Substituting ${\bf 1}_K = \frac{1}{\sigma_{\rm ul}^2}\left(\boldsymbol{\Gamma} - \boldsymbol{\Sigma}^{\Ttran}\right){\bf p}$ into \eqref{eq:optimal-power-appendix} yields 
\begin{align}
\boldsymbol{\rho} = \frac{\sigma_{\rm dl}^2}{\sigma_{\rm ul}^2}\left(\boldsymbol{\Gamma} - \boldsymbol{\Sigma}\right)^{-1}\left(\boldsymbol{\Gamma} - \boldsymbol{\Sigma}^{\Ttran}\right){\bf p}.
\end{align}
The total transmit power condition now follows from direct computation.

\bibliographystyle{IEEEtran}
\bibliography{IEEEabrv,refs,ref_book,ref_MassiveMIMO2}

\end{document}